\documentclass[review]{elsarticle}

\usepackage{graphicx,scalerel,stackengine}
\usepackage[numbers]{natbib}
\usepackage{caption}
\usepackage{subcaption}
\usepackage{lscape}
\usepackage{multirow}
\usepackage{stmaryrd}
\usepackage{fancyhdr,graphicx,amsmath,amssymb}
\usepackage[ruled,vlined]{algorithm2e}

\begin{document}

\begin{frontmatter}



\title{A Novel Hierarchical Light Field Coding Scheme Based on Hybrid Stacked Multiplicative Layers and Fourier Disparity Layers for Glasses-Free 3D Displays}






\author{Joshitha Ravishankar$^{\dagger}$}
\ead{ee19d401@smail.iitm.ac.in}

\author{Mansi Sharma$^{\dagger}$\corref{cor}}
\ead{mansisharmaiitd@gmail.com, mansisharma@ee.iitm.ac.in}

\address{Indian Institute of Technology Madras, Chennai 600036, India.}
\cortext[cor]{Corresponding author\\ $^{\quad \dagger}$These authors contributed equally to this work  }

\begin{abstract}
This paper presents a novel hierarchical coding scheme for light fields based on transmittance patterns of low-rank multiplicative layers and Fourier disparity layers. The proposed scheme identifies multiplicative layers of light field view subsets optimized using a  convolutional neural network for different scanning orders. Our approach exploits the hidden low-rank structure in the multiplicative layers obtained from the subsets of different scanning patterns. The spatial redundancies in the multiplicative layers can be efficiently removed by performing low-rank approximation at different ranks on the Krylov subspace. The intra-view and inter-view redundancies between approximated layers are further removed by HEVC encoding. Next, a Fourier disparity layer representation is constructed from the first subset of the approximated light field based on the chosen hierarchical order. Subsequent view subsets are synthesized by modeling the Fourier disparity layers that iteratively refine the representation with improved accuracy. The critical advantage of the proposed hybrid layered representation and coding scheme is that it utilizes not just spatial and temporal redundancies in light fields but efficiently exploits intrinsic similarities among neighboring sub-aperture images in both horizontal and vertical directions as specified by different predication orders. In addition, the scheme is flexible to realize a range of multiple bitrates at the decoder within a single integrated system. The compression performance of the proposed scheme is analyzed on real light fields. We achieved substantial bitrate savings and maintained good light field reconstruction quality.
\end{abstract}


\end{frontmatter}


\section{Introduction}\label{sec1}
The light field imaging and 3D display markets are growing rapidly. In recent years, we have seen autostereoscopic displays emerge as a potential alternative to stereoscopic 3D displays as it supports stereopsis and motion parallax from different viewing directions~\cite{surman2014towards, balogh2007holovizio,pseudo3d, glassesfree3d}. Parallax-based and lenticular-based 3D display technology has still not reached up to the standards to simultaneously provide direction-dependent outputs without losing out on the resolution in reconstructing dense light fields. Emerging multi-layered light field displays provide continuous motion parallax, greater depth-of-field, and wider field-of-view, which are critical for reproducing realistic 3D vision. Tensor or multi-layered displays can accurately reproduce multi-view images or light fields simultaneously with high resolution using just a few light attenuating layers \cite{li2020light, wetzstein2012tensor,sharma2016novel}.

\begin{figure}
\centering 
      \includegraphics[width=0.34\textwidth]{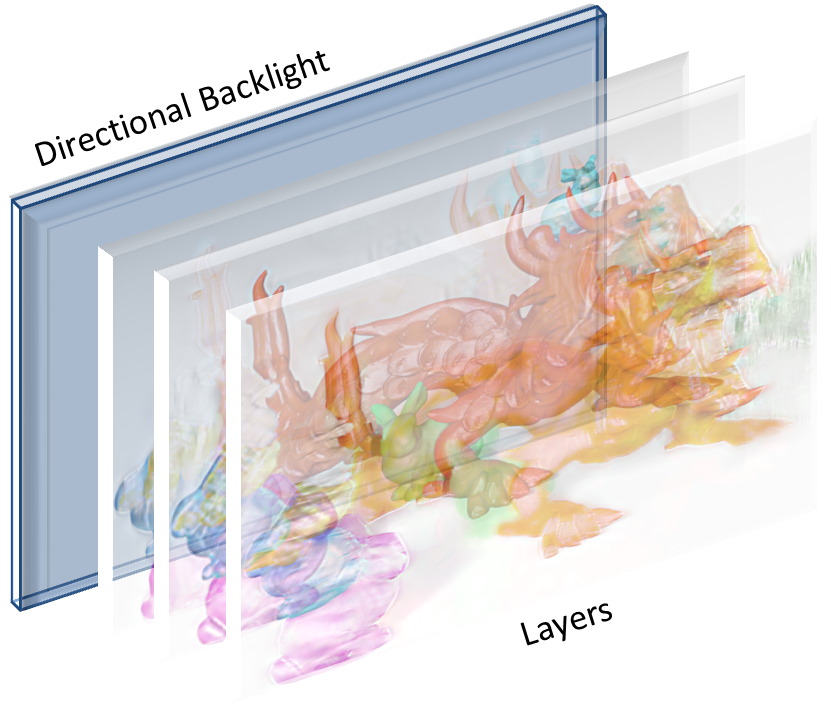}
      \caption{\footnotesize Structure of layered light field display}
      \label{fig:backlight layers}
\end{figure}

\begin{figure}[t!]
    \centering 
      \includegraphics[width=0.38\textwidth]{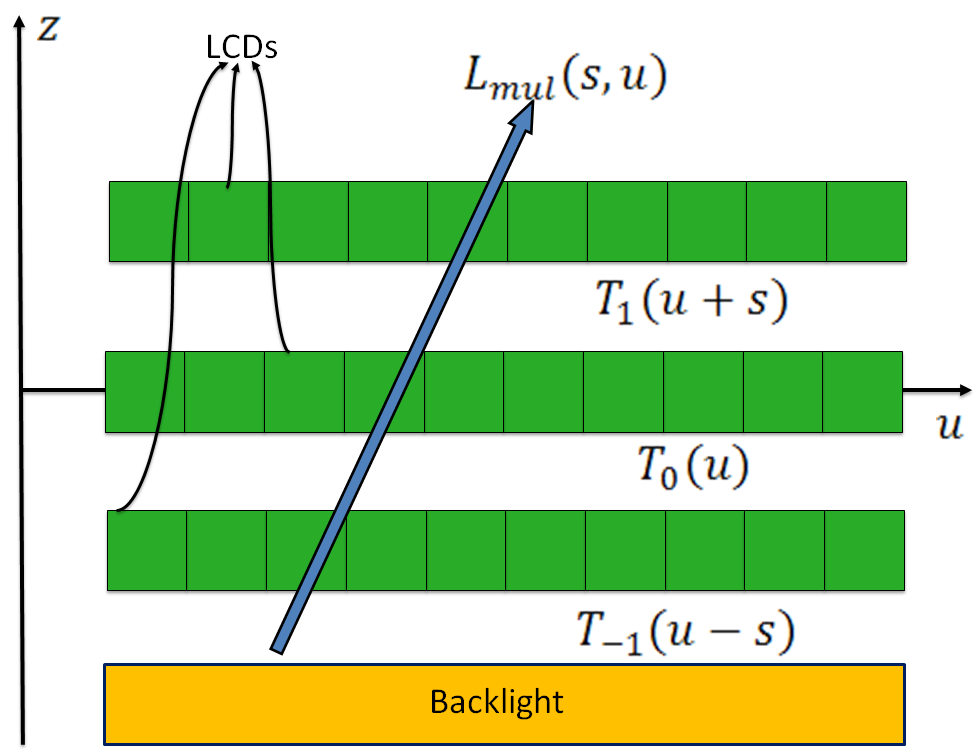}
      \caption{\footnotesize Configuration of multiplicative layers}
      \label{fig:multiplicative layers config}
\end{figure}

The typical structure of a multi-layered display is shown in Fig.~\ref{fig:backlight layers}. It consists of light-attenuating pixelized layers (\textit{e.g.}, LCD panels) stacked in front of a backlight. On each layer, the transmittance of pixels can be controlled independently by carrying out multiplicative operations. Fig.~\ref{fig:multiplicative layers config} illustrates the light rays which pass through different combinations of pixels in stacked layers depending on the viewing directions. Efficient representation and coding of light rays in multi-layered 3D displays are essential for adaptation on different auto-stereoscopic platforms. 

Existing light field coding approaches are based on raw lenslet image~\cite{RwlensletRef1_li2014efficient,RwlensletRef2_perra2016high,RwlensletRef3_li2016compression,RwlensletRef4_monteiro2017light,RwlensletRef5_liu2019content}, geometry information~\cite{RwDispRef2_jiang2017light,RwDispRef1_zhao2017light,RwEpiRef2_ahmad2020shearlet,RwEpiRef3_chen2020light}, scene content information \cite{RwCbRef1_hu2020adaptive}, disparity information~\cite{RwDispRef1_zhao2017light,RwDispRef2_jiang2017light,RwDispRef3_dib2020local}, epipolar plane image-based and multiplane image-based \cite{RwEpiRef1_vagharshakyan2017light,RwEpiRef2_ahmad2020shearlet,RwEpiRef3_chen2020light}, view prediction based learning schemes \cite{RwVsRef3_huang2019light,RwVsRef4_heriard2019light, RwDeepRef6_schiopu2019deep,RwDeepRef8_liu2021view} or methods considering light field data as a pseudo video sequence \cite{RwPsvRef1_liu2016pseudo,RwPsvRef3_ahmad2017interpreting,RwPsvRef4_ahmad2019computationally,RwPsvRef5_gu2019high}. These compression approaches are not specifically designed for tensor or multi-layered displays.

Our previous work on light field coding exploits the spatial and temporal correlations among light field multiplicative layers~\cite{ravishankar2021flexible}. We handled the inherent redundancies in light fields by approximating multiplicative layers in the image-based spatial domain. In this work, we propose a novel hierarchical coding scheme for light field compression based on a hybrid multiplicative layers \cite{maruyama2020comparison} and Fourier disparity layers representation \cite{le2019fourier}. The current approach efficiently deals with elimination of spatial, temporal and non-linear redundancies present in the light field view subsets while working in an integrated spatial and Fourier domain representation. This is a generalized coding scheme applicable to a variety of autostereoscopic displays, in particular, useful for tensor or multi-layer light field displays~\cite{surman2014towards, balogh2007holovizio,pseudo3d, glassesfree3d}. It offers much more bitrate savings and adaptability to different coding approaches and also achieves the goal of covering a range of multiple bitrates in a single unified system. Apart from supporting multi-view and layered 3D displays, our model can complement other light field coding schemes based on learning networks to support various bitrates as well~\cite{RwDeepRef3_bakir2018light, RwDeepRef4_zhao2018light,RwDeepRef5_wang2019region,RwDeepRef6_schiopu2019deep, RwDeepRef7_jia2018light,RwDeepRef8_liu2021view}.

\begin{figure*}
\centering 
 \begin{subfigure}{0.4\linewidth}
     \includegraphics[width=\linewidth]{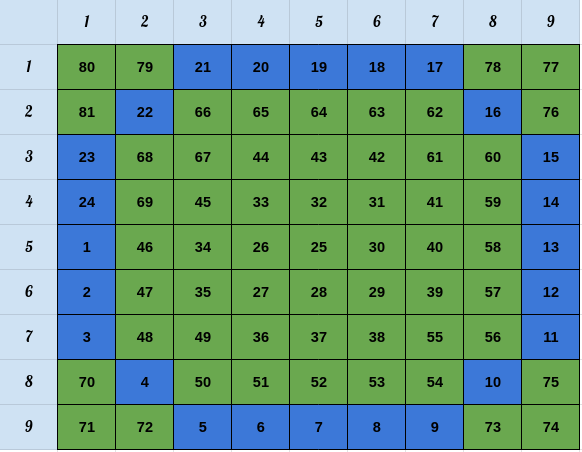}
     \caption{\footnotesize Circular-2}
 \end{subfigure}
 \begin{subfigure}{0.4\linewidth} 
     \includegraphics[width=\linewidth]{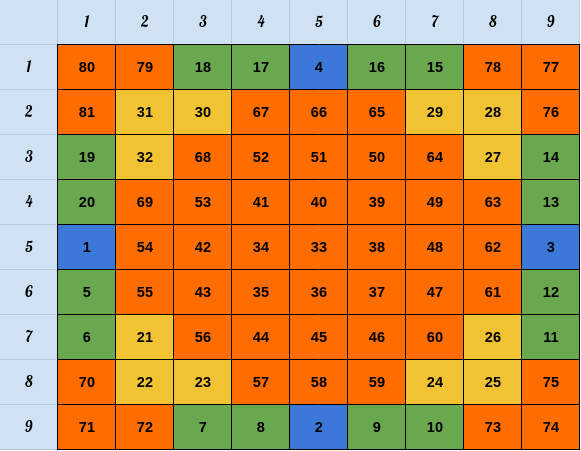}
     \caption{\footnotesize Circular-4}
 \end{subfigure} 
 \begin{subfigure}{0.4\linewidth}
     \includegraphics[width=\linewidth]{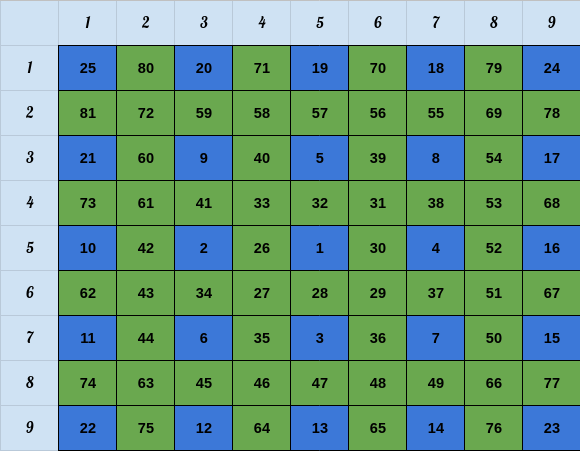}
     \caption{\footnotesize Hierarchical-2}
 \end{subfigure}
 \begin{subfigure}{0.4\linewidth} 
     \includegraphics[width=\linewidth]{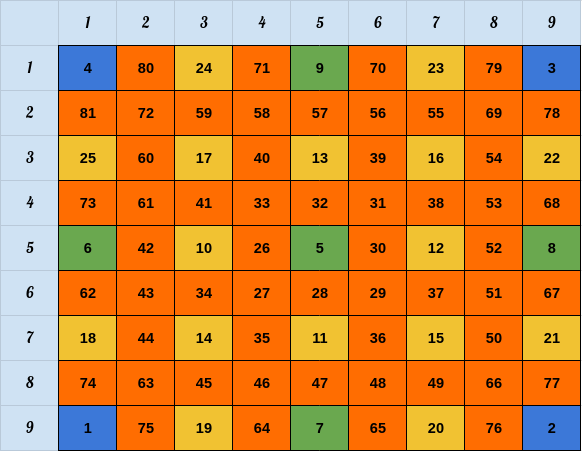}
     \caption{\footnotesize Hierarchical-4}
 \end{subfigure} 
 \caption{\footnotesize The light field view prediction orders $C_2$, $C_4$, $H_2$ \& $H_4$. Views in blue, green, yellow \& orange form the first, second, third and fourth subset respectively.}
 \label{fig:view subset orders}
\end{figure*}

In the proposed coding scheme, the input light field is divided into view subsets based on pre-defined Circular and Hierarchical prediction orders (Fig.~\ref{fig:view subset orders}). Three optimized multiplicative layers for each view subset using convolutional neural networks (CNN) are constructed. The key idea in our proposed coding scheme is to reduce the dimensionality of stacked multiplicative layers using the randomized Block-Krylov singular value decomposition (BK-SVD) \cite{musco2015randomized}. Factorization derived from BK-SVD effectively exploits the high spatial correlation between the multiplicative layers and approximates the light field subsets for varying low ranks. Encoding of these low-rank approximated subset multiplicative layers using HEVC codec \cite{sullivan2012overview} is performed to eliminate intra-view and inter-view redundancies further. Thus, our scheme approximates multiplicative layers of target light field view subsets for multiple ranks and quantization parameters (QPs) in the first stage. The view subsets are then reconstructed from the decoded layers.

In the second stage, the processing of the entire approximated light field is done in the Fourier domain, following a hierarchical coding procedure. A Fourier Disparity Layer (FDL) calibration is performed to estimate disparity values and angular coordinates of each light field view \cite{dib2019light}. These essential parameters provide additional information for the FDL construction and view prediction. They are then transmitted to the decoder as metadata. Next, we split the approximated light field into subsets as identified by four scanning or prediction orders. The first set of views are encoded and utilized for constructing the FDL representation. This FDL representation synthesizes the succeeding view subsets. The remaining correlations between the prediction residue of synthesized views and the approximated subset are further eliminated by encoding the residual signal using HEVC. The set of views obtained from decoding the residual are employed to refine the FDL representation and predict the next subset of views with improved accuracy. This hierarchical procedure continues iterating until all light field views are coded. The critical advantages of the proposed hybrid layered representation and coding scheme are:

\begin{itemize} 
\item{Our scheme efficiently exploits spatial, temporal, and non-linear redundancies between adjacent sub-aperture images in the light field structure within a single integrated framework. The scheme leverages the benefits of hybrid multiplicative layers and Fourier disparity layers representation in our hierarchical coding and prediction model for gaining superior compression efficiency without compromising reconstruction quality. In the first stage, we process light field in the spatial domain and remove the intra-view and inter-view redundancies among subset multiplicative layers. In the second stage, while working in the Fourier domain, our scheme ensures to eliminate the non-linear redundancies among adjacent views in both horizontal and vertical directions that exhibit high similarities. Experiments with various real light fields following different scanning orders demonstrate superior compression performance of our proposed model.
}
\item{The scheme is versatile to realize a range of multiple bitrates within a single integrated system trained using few convolutional neural networks. This characteristic of the proposed model complements existing light field coding systems or methods which usually support only specific bitrates during compression using multiple networks. Our coding model is adaptable to support various computational, multi-view auto-stereoscopic platforms, table-top, head-mounted displays, or mobile platforms by optimizing the bandwidth for a given target bitrate.
}
\end{itemize}

A shorter version of this work has been accepted for publication at IEEE SMC 2021~\cite{DBLP:journals/corr/abs-2104-09378}. The current journal version is an extension that elaborates on two more light field scanning patterns for view subsets. It includes new detailed results and extensive analysis of the performance of the proposed light field coding model. The rest of this article is organized into four major sections. Section~\ref{rw} describes various existing light field compression approaches and their shortcomings. The proposed layered representation and coding scheme for multi-view displays is discussed in Section~\ref{pm} in detail. We have elaborated our experiments specifying the implementation, results, and analysis in Section~\ref{ra}. Lastly, the conclusion with comprehensive findings of our proposed scheme and implications of future work are presented in Section~\ref{con}.

\section{Related Works}\label{rw}

The light field image or plenoptic image contains information about the intensity and direction of light rays in space \cite{levoy1996light,gortler1996lumigraph}. It involves a large volume of data. Thus, the storage and transmission requirements for light fields are tremendous. Efficient compression techniques leveraging the data redundancy in both the spatial and angular domains of light fields are essential. Compression or coding techniques for light fields are often broadly classified into two categories; lenslet-based or approaches based on perspective sub-aperture images. 

The direct compression using raw light field lenslet image captured by plenoptic camera exploits the spatial redundancy between the microimages~\cite{RwlensletRef1_li2014efficient,RwlensletRef2_perra2016high,RwlensletRef3_li2016compression,RwlensletRef4_monteiro2017light,RwlensletRef5_liu2019content}. This approach is usually based on the existing image/video coding standards, such as JPEG \cite{jpeg_pennebaker1992jpeg} or high efficiency video coding (HEVC) \cite{hevc_sullivan2012overview}. The unique light field structural characteristics make it harder to predict the lenslet image regions with complex textures accurately. Moreover, there are numerous microimages in the raw light field image which need careful handling. The microimages are of low resolution and require customized reshaping before being fed into an HEVC encoder. Ideally, the lenslet-based compression solutions need to transmit camera parameters for further processing, and thus increase the coding burden on the compressed data stream. These drawbacks encouraged researchers to extract the light field sub-aperture images (SAIs) from raw plenoptic images and explore compression possibilities on these views.

There are various coding methods that directly take light field SAIs as the input. They are broadly categorized as content-based compression \cite{RwCbRef1_hu2020adaptive}, disparity-based \cite{RwDispRef1_zhao2017light,RwDispRef2_jiang2017light,RwDispRef3_dib2020local}, epipolar plane image-based and multiplane image-based \cite{RwEpiRef1_vagharshakyan2017light,RwEpiRef2_ahmad2020shearlet,RwEpiRef3_chen2020light}, pseudo sequence based \cite{RwPsvRef1_liu2016pseudo,RwPsvRef2_li2017pseudo,RwPsvRef3_ahmad2017interpreting,RwPsvRef4_ahmad2019computationally,RwPsvRef5_gu2019high}, view synthesis based \cite{RwVsRef1_senoh2018efficient,RwVsRef2_huang2018view,RwVsRef3_huang2019light,RwVsRef4_heriard2019light}, and learning-based compression methods \cite{RwDeepRef3_bakir2018light, RwDeepRef4_zhao2018light,RwDeepRef5_wang2019region,RwDeepRef6_schiopu2019deep, RwDeepRef7_jia2018light,RwDeepRef8_liu2021view}.

Disparity-based compression methods approximate particular views by the weighted sum of other views \cite{RwDispRef1_zhao2017light} or apply the homography-based low-rank approximation method called HLRA \cite{RwDispRef2_jiang2017light}. This approach depends on how much the disparities across views vary and may not optimally reduce the low-rank approximation error for light fields with large baselines. Dib et al.~\cite{RwDispRef3_dib2020local} proposed a novel parametric disparity estimation method to support the low-rank approximation using super rays. They efficiently exploit redundancy across the different views compared to the homography-based alignment. A shearlet transform-based method presented in \cite{RwEpiRef2_ahmad2020shearlet} categorizes the SAIs into key and decimated views. The scheme performs well under low bitrates. The decimated views are predicted from the compressed key views and a residual bitstream. Chen et al. \cite{RwEpiRef3_chen2020light} used multiplane representation and strongly reduced the calculation burden on the decoder.

\begin{figure*}
    \centering 
    \begin{subfigure}{\linewidth}
      \includegraphics[width=\linewidth]{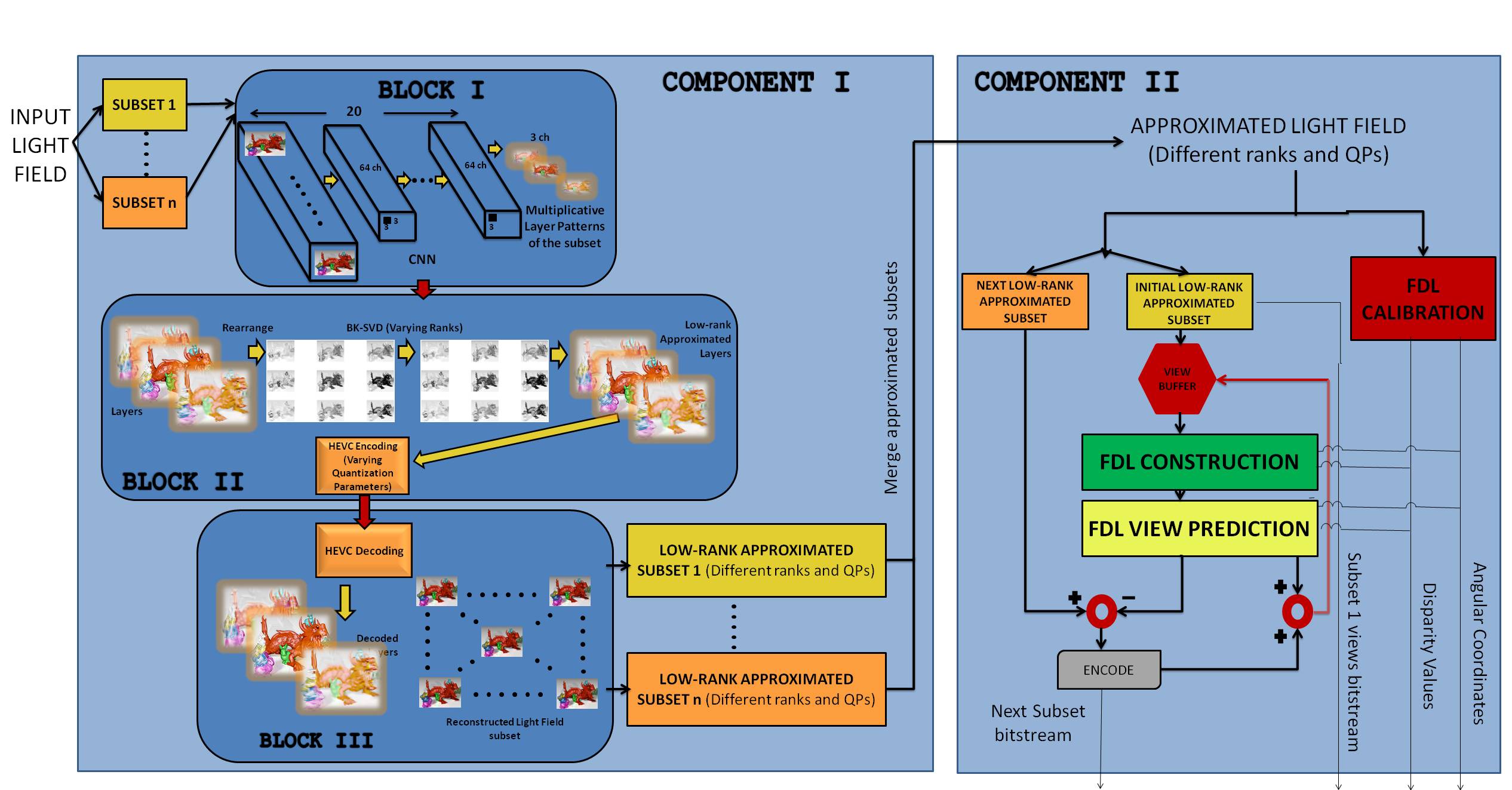}
      \caption{\footnotesize}
      \label{fig:workflow_encoding}
    \end{subfigure}
     \begin{subfigure}{0.6\linewidth}
      \includegraphics[width=\linewidth]{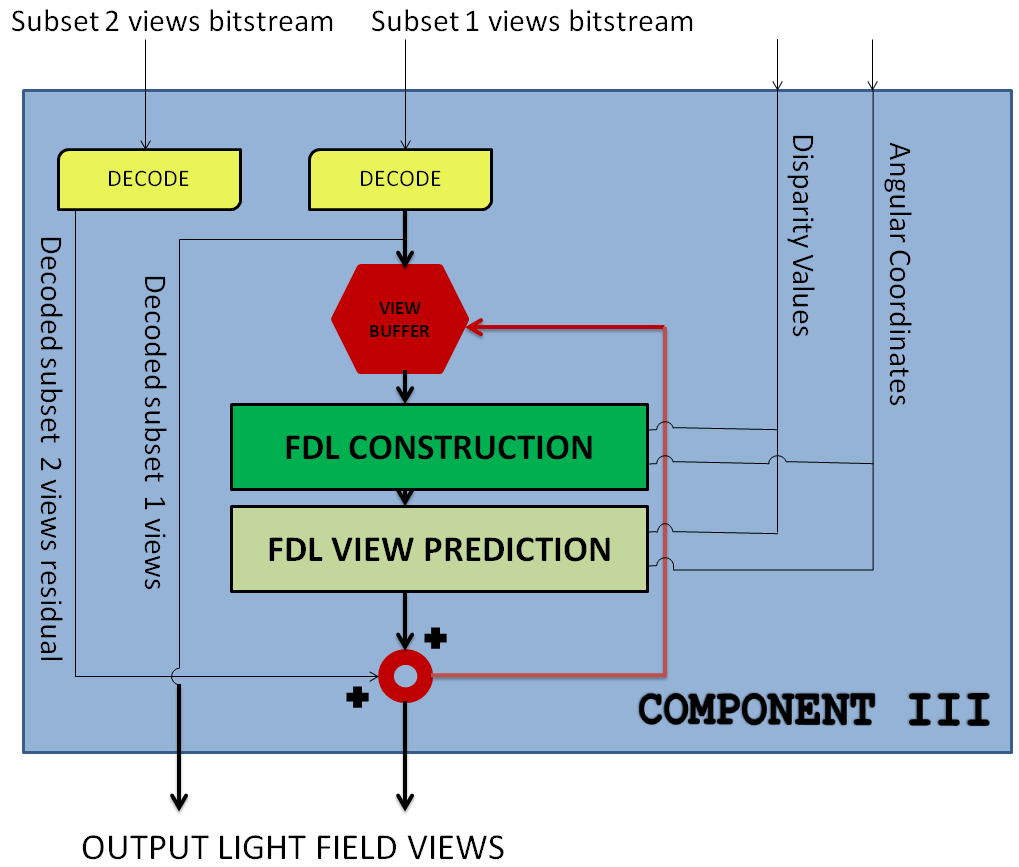}
      \caption{\footnotesize}
      \label{fig:workflow_decoding}
    \end{subfigure}
\caption{\footnotesize Three components of proposed light field coding scheme. (a) Overview of the encoding scheme (b) Overview of the decoding scheme. }
\label{fig:main workflow}
\end{figure*}

The views are reordered as pseudo sequences in predictive compression methods. Liu et al. \cite{RwPsvRef1_liu2016pseudo} in their symmetric 2D hierarchical order, compresses the central view first as the I-frame (intra frame) followed by the remaining views as P-frames. Their prediction structure ensures inter-view prediction from adjacent views.  Another proposal by Li et al. \cite{RwPsvRef2_li2017pseudo} involves dividing all the light field views into four quadrants and adopting a hierarchical coding structure within each quadrant. Ahmad et al. \cite{RwPsvRef3_ahmad2017interpreting} utilize the multi-view extension of HEVC (MV-HEVC) to compress the light field in the form of a multi-view sequence. They interpret each row of the sub-aperture views as a single view of a multi-view video sequence and propose a two-dimensional prediction and rate allocation scheme. The extension of their work \cite{RwPsvRef4_ahmad2019computationally} uses hierarchical levels, where views belonging to higher levels are assigned with better quality. The higher-level views are used as references to predict the lower-level views. Synthesized virtual reference frames are generated from Adaptive Separable Convolution Network (ASCN) in another SAIs based technique~\cite{RwPsvRef5_gu2019high}. Such frames are considered as extra reference candidates in a hierarchical coding structure for MV-HEVC to further exploit intrinsic similarities in light field images.

A block-basis estimation of views from translated reference views is proposed in \cite{RwVsRef4_heriard2019light}. The residuals of estimated views are also transmitted to the decoder along with the rest of the view estimation parameters. Other view synthesis compression schemes estimate image depth maps from a subset of reference light field views. The Multi-view Video plus Depth (MVD) structure is adopted for depth image-based rendering to synthesize the intermediate SAIs \cite{RwVsRef2_huang2018view}. A pair of steps to generate noise-refined depth maps for selected perspective views is elaborated in \cite{RwVsRef3_huang2019light}.

Bakir et al.~\cite{RwDeepRef3_bakir2018light} presented a deep learning-based scheme on the decoder side to improve the reconstruction quality of sub-aperture images. Similarly, Zhao et al. \cite{RwDeepRef4_zhao2018light} encoded only sparsely sampled SAIs, while the remaining SAIs are synthesized using a CNN from the decoded sampled SAIs as priors. However, these methods require large-scale and diverse training samples, and high quality of reconstructed views is only obtained if at least half the SAIs are taken as reference. Schiopu et al. \cite{RwDeepRef6_schiopu2019deep} proposed a novel network that synthesizes the entire light field image as an array of synthesized macro pixels in one step. Wang et al.~\cite{RwDeepRef5_wang2019region} identify a region of interest (ROI), a complex non-ROI, and a smooth non-ROI to compress light field videos framewise. A generative adversarial network (GAN) is proposed in \cite{RwDeepRef7_jia2018light} for unsampled SAIs generation. Liu et al. \cite{RwDeepRef8_liu2021view} also adopt a GAN framework to boost the light field compression. They use an image group-based sampling method to reduce more SAI redundancy and maintain the reconstructed SAI quality. A perceptual quality-based loss function is also proposed by considering the PSNR of synthetic SAIs and adversarial loss. The above-mentioned coding techniques are not explicitly designed for the representation used in multi-layer-based light field displays. They also usually train a system (or network) to support only specific bitrates during the compression.


\section{Proposed Methodology}\label{pm}

The complete workflow of our proposed representation and coding scheme with three main components is illustrated in Fig.~\ref{fig:main workflow}. In COMPONENT I, input light field images are divided into view subsets depending on the specific prediction orders. To efficiently exploit the intrinsic redundancies in light field data, the proposed scheme constructs multiplicative layers from each view subset by employing a CNN. BLOCK I represents a CNN that converts the light field views of the input subsets into three multiplicative layers \cite{maruyama2020comparison}. In BLOCK II, we removed the intrinsic redundancy present in subset views by exploiting the hidden low-rank structure of multiplicative layers on a Krylov subspace \cite{musco2015randomized}. The low-rank approximation of each subset multiplicative layers is performed using Block-Krylov singular value decomposition (BK-SVD) by choosing various ranks~\cite{musco2015randomized}. The intra-frame and inter-frame redundancies are further eliminated by encoding the approximated layers with HEVC~\cite{sullivan2012overview}. We reconstruct the approximated views of each subset from their respective decoded multiplicative layers in BLOCK III. At the end of COMPONENT I of our scheme, we obtain approximated light field data at various ranks and quantization parameters.

In the next second phase, approximated subsets are used to construct Fourier disparity layer (FDL) representation of light fields~\cite{le2019fourier}. The processing of approximated light field in COMPONENT II of our scheme is shown in Fig.~\ref{fig:workflow_encoding}. There exist non-linear correlations between neighboring sub-aperture views in both horizontal and vertical directions in the light field structure. We particularly target these redundancies between adjacent light field views by processing in the Fourier domain as specified by different scanning or predication orders. The light field is then iteratively reconstructed by the FDL representation in a hierarchical fashion. We find the angular coordinates and disparity values of each view of the low-rank approximated light field (at different ranks and QPs) through FDL calibration and directly transmit them to the decoder (COMPONENT III) as metadata~\cite{dib2019light}. The approximated light field is divided into subsets specified by four different prediction orders. The initial subset is used in construction of the FDL representation, which is further employed to synthesize the subsequent subsets of views. The correlations in prediction residue are removed, and a more accurate FDL representation is constructed from previously encoded subsets. Thus, we iteratively refine the FDL representation in COMPONENT II until all the approximated light field views are encoded. The decoding scheme is depicted in COMPONENT III (Fig.~\ref{fig:workflow_decoding}). Here, angular coordinates and disparity values of the low-rank approximated light field, along with the encoded bitstreams of approximated subsets are utilized for the final light field reconstruction.

\begin{figure}
    \centering
    \includegraphics[scale=0.17]{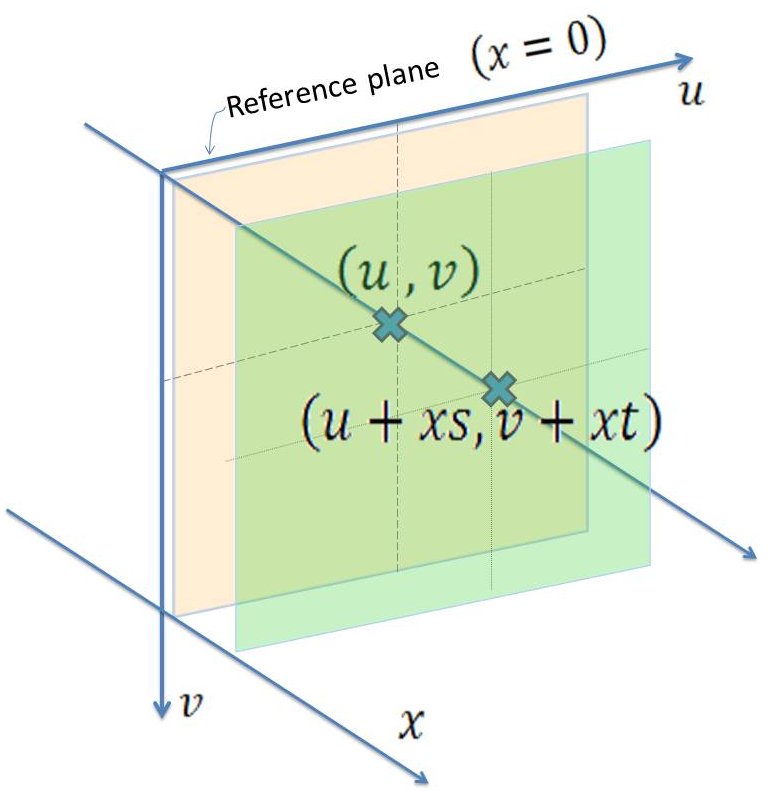}
    \caption{\footnotesize The light ray is parameterized by point of intersection with the $(u,v)$ plane and the $(s,t)$ plane located at a depth $x$.}
    \label{fig:lf planes}
\end{figure}


\subsection{Approximation of light field at different ranks}

The input light field is divided into view subsets in COMPONENT I, depending on the specified scanning orders, and a low-rank approximated representation of the subset multiplicative layers is obtained in the Krylov subspace. BLOCK I to III in Fig.~\ref{fig:workflow_encoding} illustrate the three sub-blocks involved in this step. The details of each step are described in the following sub-sections.


\subsubsection{View Subsets of Light Field}\label{sec:view_subsets}

The proposed scheme divides the light field into different view subsets based on predefined scanning orders. We adopt a hierarchical pattern configuration and a circular view prediction order described in the work of Dib et al.~\cite{dib2019light}. The four chosen patterns are Circular-2 ($C_2$), Circular-4 ($C_4$), Hierarchical-2 ($H_2$) and Hierarchical-4 ($H_4$). For a 9$\times$9 light field, the $C_2$ and $H_2$ patterns contain two view subsets and $C_4$ and $H_4$ patterns have four subsets. The exact coding orders of each subset of these four chosen scanning orders are shown in Fig.~\ref{fig:view subset orders}. In all subsets of these patterns, the light field views form a circle that spiral out from the center. Generally, the corner views of light fields are of lower quality, and thus we choose to form a circle rather than a square while scanning the views. Our proposed workflow begins with partitioning of the input light field into subsets based on $C_2$, $C_4$, $H_2$ or $H_4$ patterns (Fig.~\ref{fig:main workflow}).


\begin{figure*}
    \centering 
    \begin{subfigure}{0.29\textwidth}
      \includegraphics[width=\linewidth]{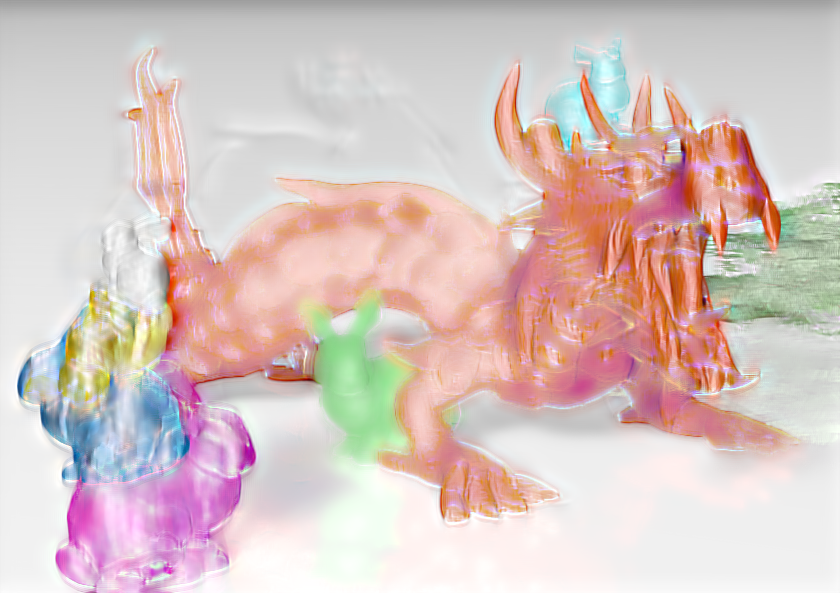}
      \caption{\footnotesize Layer -1}
    \end{subfigure}
     \begin{subfigure}{0.29\textwidth}
      \includegraphics[width=\linewidth]{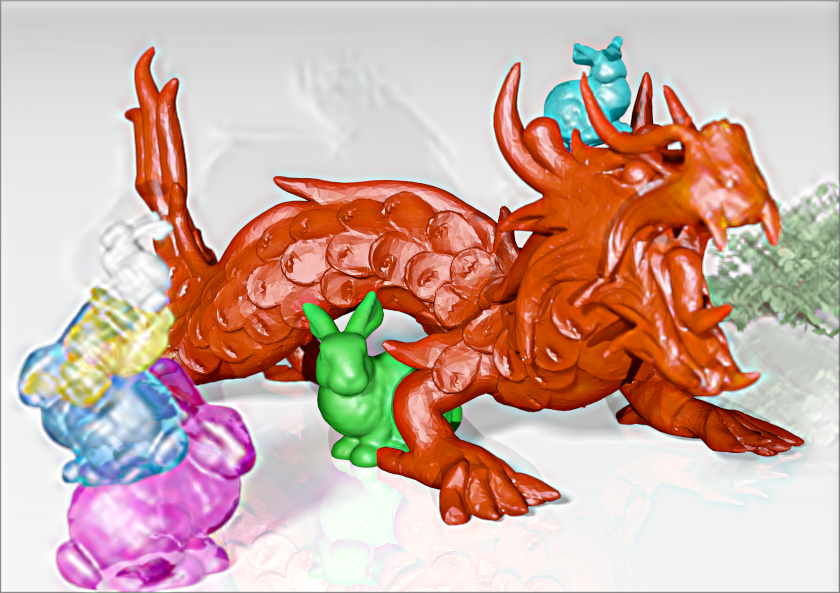}
      \caption{\footnotesize Layer 0}
    \end{subfigure}
     \begin{subfigure}{0.29\textwidth}
      \includegraphics[width=\linewidth]{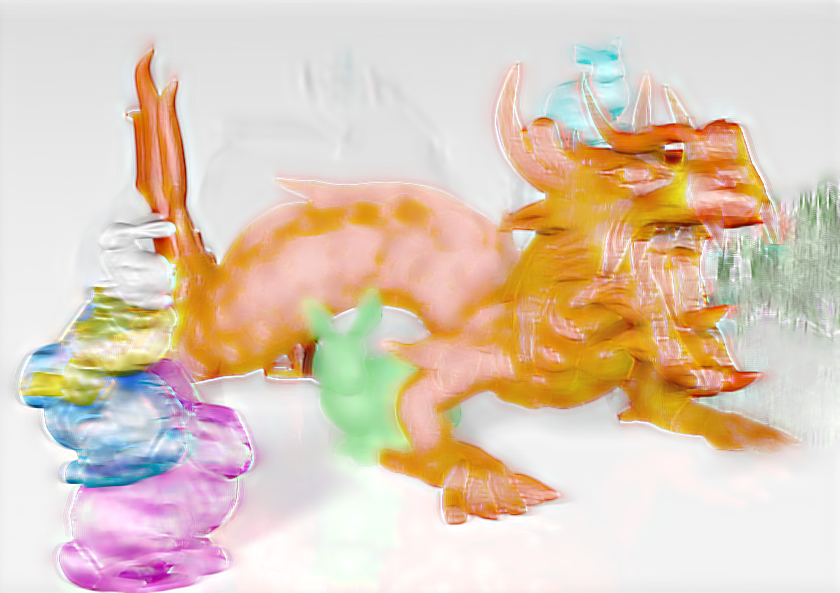}
      \caption{\footnotesize Layer +1}
    \end{subfigure}
\caption{\footnotesize The three optimal multiplicative layers produced by the CNN for  the \textit{Dragon and Bunnies} light field~\cite{mitdragon}. }
\label{fig:dragon_layers}
\end{figure*}

\subsubsection{Light Field Views to Stacked Multiplicative Layers} 

The light field  $L(s,t,u,v)$ characterizes radiance along light rays as a 4D function~\cite{levoy1996light,gortler1996lumigraph}. In $L(s,t,u,v)$, the parameters $(s,t)$ specify the spatial coordinates  and $(u,v)$ denote the angular coordinates that represent intersection of the rays with two parallel planes (Fig.~\ref{fig:lf planes}). A light field can be interpreted as a set of directional views and can be produced using stacked layers that carry out multiplicative light ray operations~\cite{maruyama2020comparison}. Such multiplicative layers could be realized with few light-attenuating panels stacked in equally spaced intervals in front of a backlight (Fig.~\ref{fig:backlight layers}).

The normalized intensity of a light ray emitted from light field display can be described as 
\begin{equation}
\centering
    L_{m}(s,t,u,v)= \prod_{x \in X}T_{x}(u+xs,v+xt)
    \label{eq:mult layers}
\end{equation}
where, $T_{x}(u,v)$ is the transmittance of the layer present at disparity $x$ and $X$ denotes the set of disparities among directional views. We determined the multiplicative layers for each subset of different view scanning order. The experiments are performed by considering the light field display composed of three layers located at $X = \{ -1, 0, 1\}$.

The three multiplicative layers obtained from each subset are optimized using a data-driven CNN based approach. The network architecture is depicted in Fig.~\ref{fig:workflow_encoding}. The objective is to optimize the layer patterns. Mathematically, it is expressed as 
\begin{equation}
    \underset{T_{x}|x \in X}{\mathrm{argmin}} \sum_{s,t,u,v}\left \| L(s,t,u,v)-L_{m}(s,t,u,v) \right \|^{2}
    \label{eqn:opteqn}
\end{equation}
The optimized multiplicative layers obtained for $5 \times 5$ \textit{Dragon and Bunnies} light field is shown in Fig.~\ref{fig:dragon_layers}.

\subsubsection{Low-Rank Representation of Subset Multiplicative Layers on Krylov Subspace} \label{sec:bksvd}

The randomized Block-Krylov SVD method introduced by Cameron and Christopher can optimally achieve the low-rank approximation of a matrix within $(1 + \epsilon)$ of optimal for spectral norm error~\cite{musco2015randomized}. The algorithm quickly converges in $\tilde{O}(\frac{1}{\sqrt{\epsilon}})$ iterations for any matrix. In our proposed scheme, the optimized multiplicative layers are compactly represented on a Krylov subspace in order to remove the intrinsic redundancy.

We denote each multiplicative layer pattern obtained from the CNN as $T_{x} \in \mathbb{R}^{m \times n \times 3}$, where $ x \in \{ -1, 0, 1\}$. The red, green, and blue colour channels of the layer $x$ are denoted as $T^{r}_{x}$, $T^{g}_{x}$, and $T^{b}_{x}$ respectively. We construct three matrices $A^{ch} \in \mathbb{R}^{3m \times n}$, $ch \in \{ r, g, b \}$ as 
\begin{equation}
\centering
A^{ch}= \left( \begin{array}{cc}
    \: \:T^{ch}_{-1} \\ 
    T^{ch}_{0} \\ 
    T^{ch}_{1} \end{array} \right)
\end{equation}
The BK-SVD low-rank approximation in a Krylov subspace for each $A^{ch}$ can remove the intrinsic redundancies in multiplicative layers of the subsets. For simplicity, we will denote $A^{ch}$ as $A$ henceforth. To approximate $A$, the objective is to  achieve a subspace that closely captures the variance of $A$'s top singular vectors and avoid the gap dependence in singular values. 
The spectral norm low-rank approximation error of $A$ is 
\begin{equation}
    \left \| A - D D^{T} A  \right \|_{2} \leq \left ( 1 + \epsilon  \right ) \left \|A - A_{k}  \right \|_{2}
    \label{eq:spectral norm error}
\end{equation}
Only top $k$ singular vectors of $A$ are used in its rank-$k$ approximation $A_{k}$. If $D$ is a rank-$k$ matrix with orthonormal columns, the spectral norm guarantee ensures that $D D^{T} A$ recovers $A$ up to a threshold $\epsilon$.

Block Krylov iteration is performed working with the Krylov subspace
\begin{equation}
K =[\Pi \hspace{8pt} A\Pi  \hspace{8pt} A^{2} \Pi \hspace{8pt} A^{3}\Pi \cdot \cdot \cdot A^{q}\Pi ]
\label{eq: Kspace} 
\end{equation}

As analyzed in~\cite{musco2015randomized}, we choose to work on low degree polynomials that allow  faster computation in fewer powers of $A$, and thus enabling convergence of the BK-SVD algorithm in fewer iterations. From subspace $K$, we construct $p_{q}(A)\Pi$ for any polynomial $p_{q}(\cdot)$ of degree $q$, where $\Pi \sim N (0,1)^{d \times k}$. The approximation of $A$ done by projecting it onto the span of $p_{q}(A)\Pi$ atleast matches the best $k$ rank approximation of $A$ lying in the span of Krylov space $K$~\cite{musco2015randomized}. Further, we orthonormalize columns of $K$ to obtain $Q \in \mathbb{R}^{c \times qk}$ using QR decomposition method~\cite{gu1996efficient}. After computing  SVD of matrix $S = Q^{T} B B^{T} Q$, we found matrix $\bar{U}_{k}$ containing the top $k$ singular vectors of $S$. Thus, the rank-$k$ approximation of $A$ is matrix $D$, which is computed as 
\begin{equation}
    D = Q \bar{U}_{k}
\end{equation}

Consequently, the rank-$k$ block Krylov approximation of matrices $A^{r}$ , $A^{g}$ and $A^{b}$ are ${D}^{r}$, ${D}^{g}$, and ${D}^{b}$ respectively. Matrix $D^{ch} \in \mathbb{R}^{y \times z}$ for every colour channel $ch$.
To obtain the approximated layers $\bar{T}_{x}$, we extract colour channels from the approximated ${D}^{ch}$ matrices by sectioning out the rows uniformly as 
\begin{flalign*}
    & \bar{T}^{ch}_{-1} = {D}^{ch}[1 : y \: \:,\: \: 1 : z \: \:] \\
    & \bar{T}^{ch}_{0} = {D}^{ch}[y+1 :2y \:,\: \: 1 : z \: \:]   \\
    & \bar{T}^{ch}_{1} = {D}^{ch}[2y+1 : 3y \:,\: \: 1 : z \: \:] 
\end{flalign*}
The red, green, and blue colour channels are combined to form each approximated layer $\bar{T}_{-1}$, $\bar{T}_{0}$, and $\bar{T}_{1}$. Thus, factorization derived from BK-SVD exploits the spatial correlation in multiplicative layers of the subset views for varying low ranks. The three block Krylov approximated layers for each subset are subsequently encoded into a bitstream using HEVC for various QPs to further eliminate intra and inter layer redundancies in the low-rank approximated representation. The low-rank representation and coding of stacked multiplicative layers on Krylov subspace is depicted in BLOCK II of Fig.~\ref{fig:workflow_encoding}.

\begin{figure*}
    \centering 
    \begin{subfigure}{0.3\textwidth}
      \includegraphics[width=\linewidth]{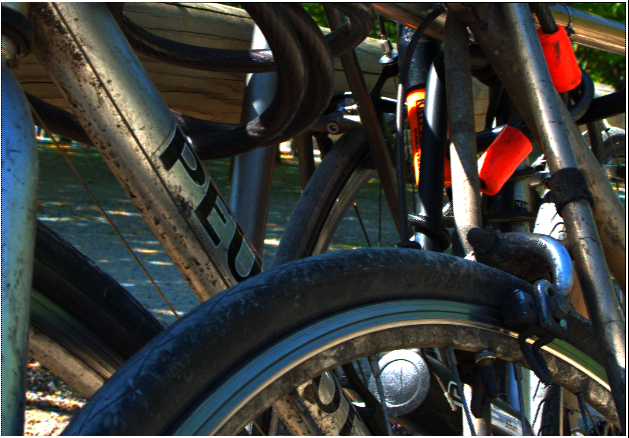}
      \caption{\footnotesize Bikes}
    \end{subfigure}
     \begin{subfigure}{0.3\textwidth}
      \includegraphics[width=\linewidth]{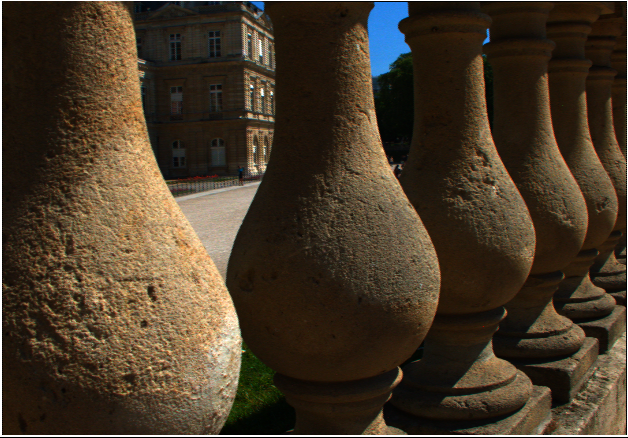}
      \caption{\footnotesize Stone pillars outside}
    \end{subfigure}
     \begin{subfigure}{0.3\textwidth}
      \includegraphics[width=\linewidth]{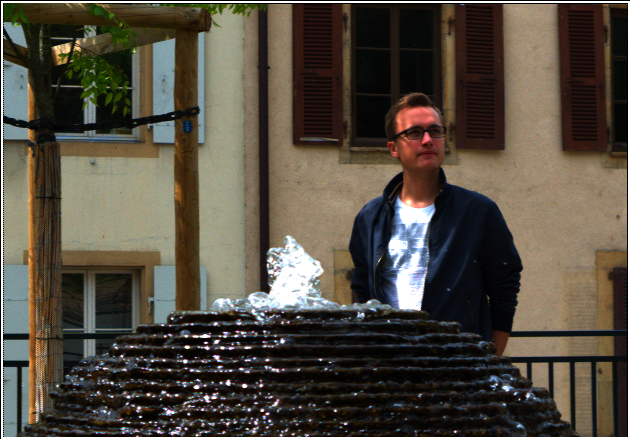}
      \caption{\footnotesize Fountain \& Vincent 2}
    \end{subfigure}
\caption{\footnotesize Central views of the three datasets. }
\label{fig:orglfs}
\end{figure*}

\subsubsection{Decoding and Reconstruction of the Light Field Subsets}

The decoding procedure of compressed layers and the reconstruction of light field subsets is shown in BLOCK III of Fig.~\ref{fig:workflow_encoding}. We decoded three multiplicative layers from the bitstream as $\grave{T}_{x}$, $x \in \{ -1, 0, 1 \}$. We consider the integers $s^{*}, t^{*}$, which vary depending on the number of views in each subset. The view $I_{(s^{*},t^{*})}$ is obtained by translating the decoded layers to $\widehat{\textit{T}}_{x}$. This step is followed by an element-wise product of colour channels of the translated layers. For a particular view $(s^{*},t^{*})$, the translation of every $x^{th}$ layer $\grave{T}_{x}$, to $\widehat{T}_{x}$ is carried out as  
\begin{equation}
     \widehat{T}_{x(s^{*},t^{*})}(u,v) = \grave{T}_{x}(u+xs^{*}, v+xt^{*})
\end{equation}
Thus, the three translated layers for every viewpoint $(s^{*},t^{*})$ is computed as  
\begin{flalign*}
    & \widehat{\textit{T}}_{-1(s^{*},t^{*})}(u,v) = \grave{T}_{-1}(u-s^{*}, v-t^{*}) \\
    & \widehat{\textit{T}}_{0 (s^{*},t^{*})}(u,v)\:\:\: = \grave{T}_{0}\:(u, v)  \\
    & \widehat{\textit{T}}_{1 (s^{*},t^{*})}(u,v)\:\:\: = \grave{T}_{1}(u+s^{*}, v+t^{*})
\end{flalign*}
An element-wise product of each colour channel $ch \in \{r, g, b \}$ of the translated layers give the corresponding colour channel of the subset view.
\begin{equation}
I^{ch}_{(s^{*},t^{*})}= \widehat{\textit{M}}^{ch}_{-1(s^{*},t^{*})} \odot  \widehat{\textit{M}}^{ch}_{0(s^{*},t^{*})} \odot \widehat{\textit{M}}^{ch}_{1(s^{*},t^{*})}
\end{equation}
The combined red, green and blue colour channels output the reconstructed light field subset at the viewpoint $(s^{*},t^{*})$ as $I_{(s^{*},t^{*})}$. At last, we merged the view subsets according to $C_2$, $C_4$, $H_2$ or $H_4$ patterns.

Thus, the spatial correlation in multiplicative layers of the subset views is exploited for different low ranks in COMPONENT I of the proposed scheme. Intra and inter-layer redundancies in the low-rank approximated representation are also well removed. We further compressed the light field by eliminating intrinsic similarities caused by non-linear correlations among neighboring views in horizontal and vertical directions as specified by $C_2$, $C_4$, $H_2$ and $H_4$ patterns using the Fourier Disparity Layers (FDL) representation. The following section describes the processing of approximated light fields in the Fourier domain.


\subsection{Fourier Disparity Layers Representation \& Light Field Processing}

The Fourier Disparity Layers representation \cite{le2019fourier} samples the input BK-SVD approximated light field in the disparity dimension by decomposing it as a discrete sum of layers. The layers are constructed from the approximated light field sub-aperture views through a regularized least square regression performed independently at each spatial frequency in the Fourier domain. The FDL representation has been shown to be effective for numerous light field processing applications \cite{le2019fourier,dib2019light, le2020hierarchical, le2020high}. We have summarised the use of FDL and encoding of low-rank approximated light field as COMPONENT II in Fig.~\ref{fig:workflow_encoding}. The corresponding decoding and reconstruction of the light field subsets are illustrated as COMPONENT III in Fig.~\ref{fig:workflow_decoding}. The Fourier Disparity Layer calibration, subset view synthesis, and prediction are described in the following subsections.


\subsubsection{FDL Calibration}

Without loss of generality, we have considered one spatial coordinate $s$ and one angular coordinate $u$ of 4D light field to present the notations in a simple manner. The approximated light field view $L_{u_{o}}$ at angular position $u_{o}$ can be defined as $L_{u_{o}}(s)=L(s,u_{o})$. We can obtain Fourier coefficients of the $j^{th}$ input light field view using $n$ disparity values $\{d_{k}\}_{k \in  \llbracket 1,n  \rrbracket }$~\cite{dib2019light}. The Fourier transform of $L_{u_{o}}$ at spatial frequency $f_{s}$ is
\begin{equation}\label{eq:ft}
    \hat{L}_{u_{o}}(f_{s})=\sum_{k} e^{+2i\pi u_{o}d_{k}f_{s}}\hat{L}^{k}(f_{s})
\end{equation}
Here, the Fourier coefficients of the disparity layers for a particular frequency $f_{s}$ is defined by 
\begin{equation}
    \hat{L}^{k}(f_{s})= \int_{\Omega_{k}}e^{-2i\pi s f_{s}}L(s,0)dx
\end{equation}
Each of such Fourier coefficients can be understood as the Fourier transform of the central light field view (as $u_{o}=0$ by just considering the spatial region $\Omega_{k}$ of disparity $d_{k}$.

The angular coordinates $u_j$ of the input views and the disparity values of the layers $d_{k}$ are estimated by minimizing over all frequency components $f^{q}_{s}$, $q \in \llbracket 1,Q\rrbracket$, where $Q$ is the number of pixels in each input image~\cite{le2019fourier}. By computing Fourier transforms of all $m$ approximated light field views as $\hat{L}_{u_{j}} (j \in \llbracket 1,m\rrbracket)$, the FDL representation is learnt by solving a linear regression problem for each frequency $f_{s}$.

The optimization problem is formulated as $\textbf{Ax} = \textbf{b}$ with Tikhonov regularization, where $\textbf{A} \in \mathbb{R}^{m \times n}$, $\textbf{x} \in \mathbb{R}^{n \times 1}$ and $\textbf{b} \in \mathbb{R}^{m \times 1}$. Elements of matrix \textbf{A} are $\textbf{A}_{jk} = e^{+2i\pi u_{j}d_{k}f_{s}}$ and \textbf{x} contains the Fourier coefficients of disparity layers $\textbf{x}_{k} = \hat{L}^{k}(f_{s})$. The vector $\textbf{b}$ contains the Fourier coefficients of $j^{th}$ input view, $\textbf{b}_{j} = \hat{L}_{u_{j}}(f_{s})$. The solution of the optimization problem results in the disparity values $d_k$ and view positions $u_j$ that are passed as metadata information to the decoder in COMPONENT III.

\begin{figure*}
\centering
\begin{subfigure}{0.13\textwidth}
  \includegraphics[width=\linewidth]{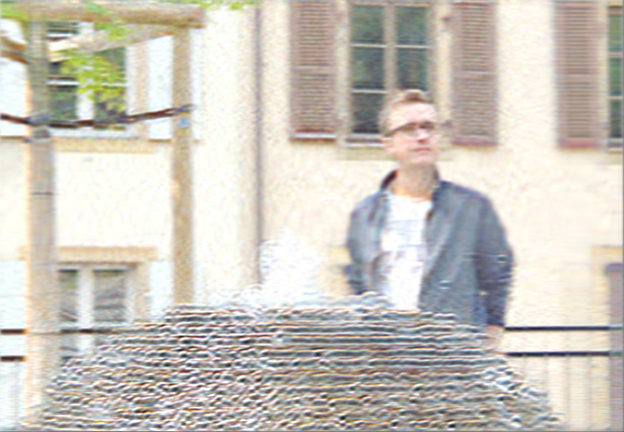}
\end{subfigure} 
\begin{subfigure}{0.13\textwidth}
  \includegraphics[width=\linewidth]{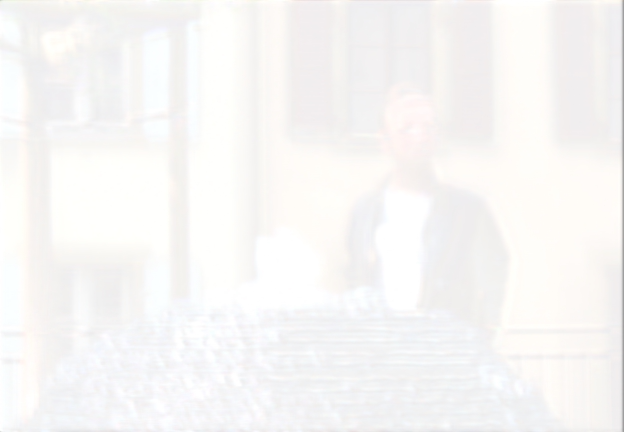}
\end{subfigure} 
\begin{subfigure}{0.13\textwidth}
  \includegraphics[width=\linewidth]{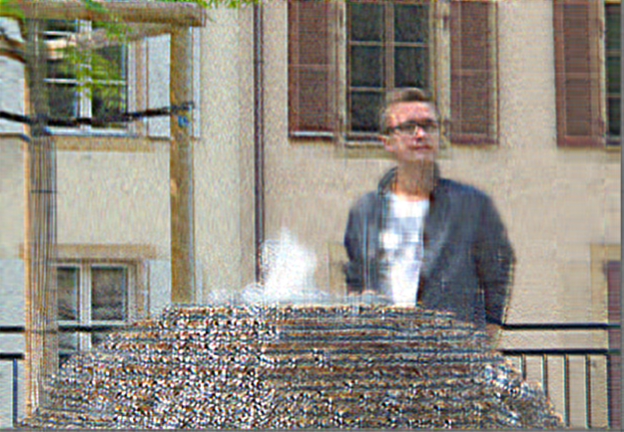}
\end{subfigure}
\begin{subfigure}{0.13\textwidth}
  \includegraphics[width=\linewidth]{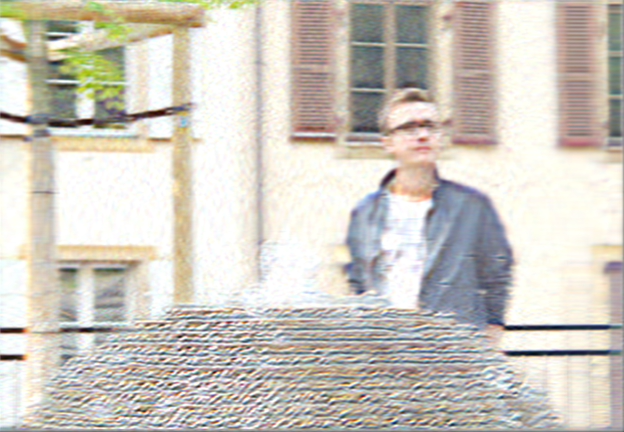}
\end{subfigure} 
\begin{subfigure}{0.13\textwidth}
  \includegraphics[width=\linewidth]{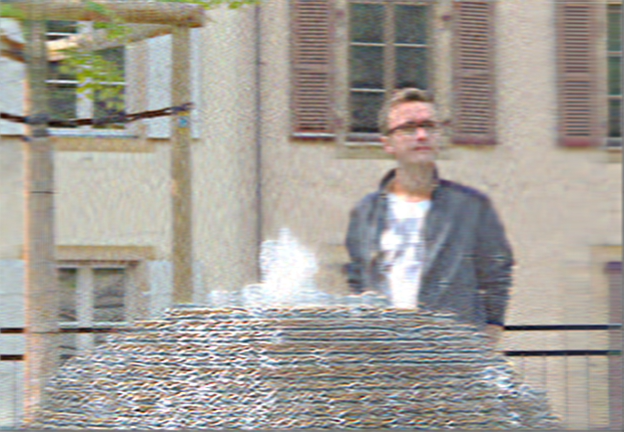}
\end{subfigure} 
\begin{subfigure}{0.13\textwidth}
  \includegraphics[width=\linewidth]{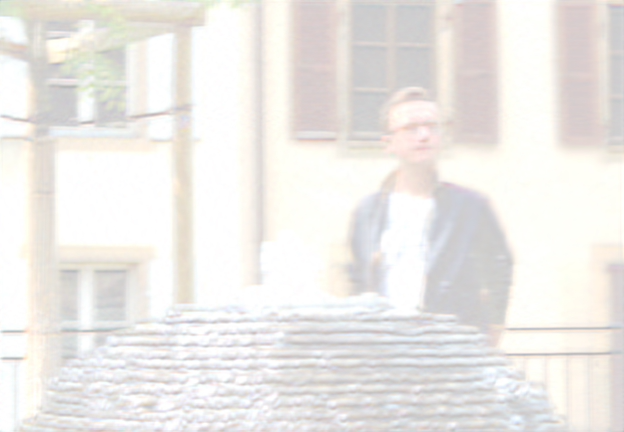}
\end{subfigure}

\begin{subfigure}{0.13\textwidth}
  \includegraphics[width=\linewidth]{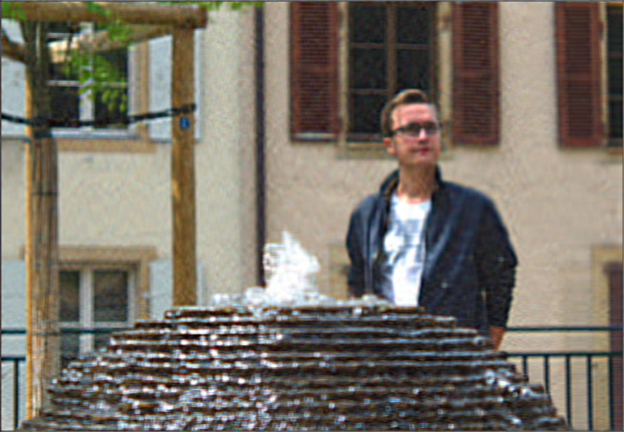}
\end{subfigure} 
\begin{subfigure}{0.13\textwidth}
  \includegraphics[width=\linewidth]{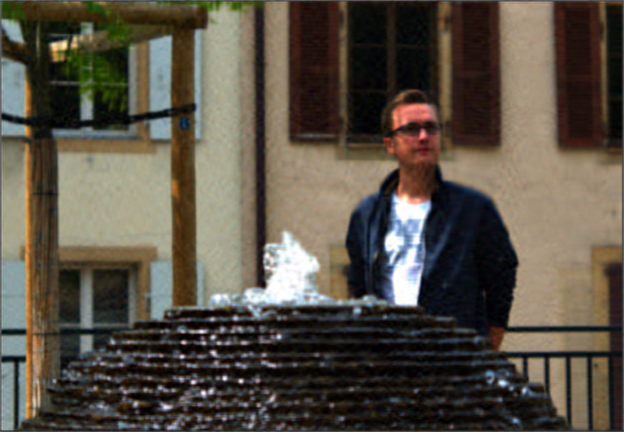}
\end{subfigure} 
\begin{subfigure}{0.13\textwidth}
  \includegraphics[width=\linewidth]{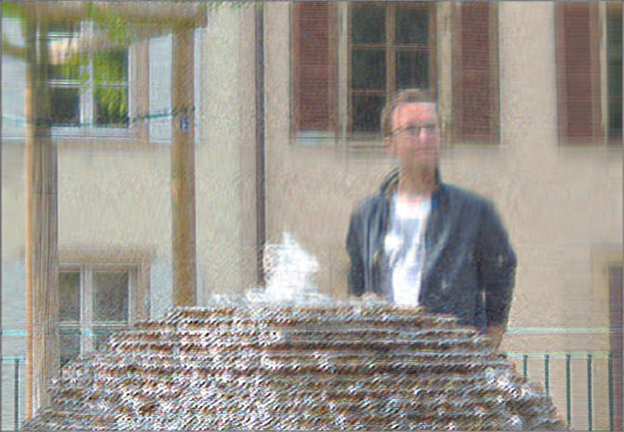}
\end{subfigure}
\begin{subfigure}{0.13\textwidth}
  \includegraphics[width=\linewidth]{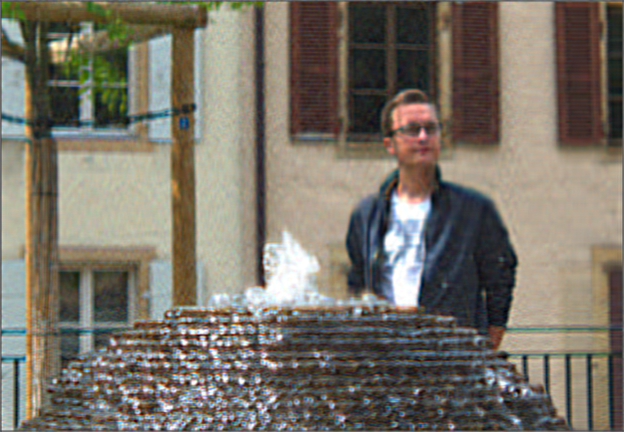}
\end{subfigure} 
\begin{subfigure}{0.13\textwidth}
  \includegraphics[width=\linewidth]{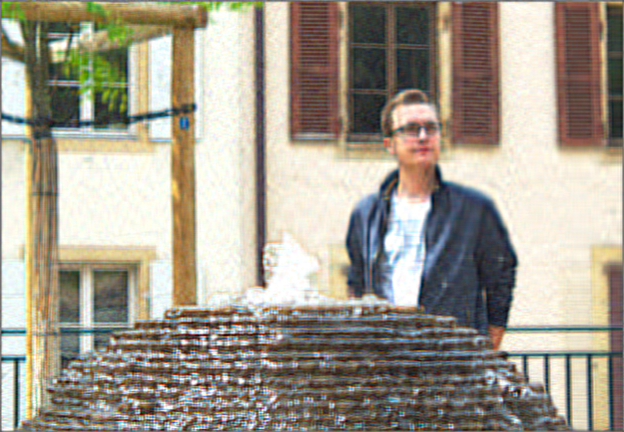}
\end{subfigure} 
\begin{subfigure}{0.13\textwidth}
  \includegraphics[width=\linewidth]{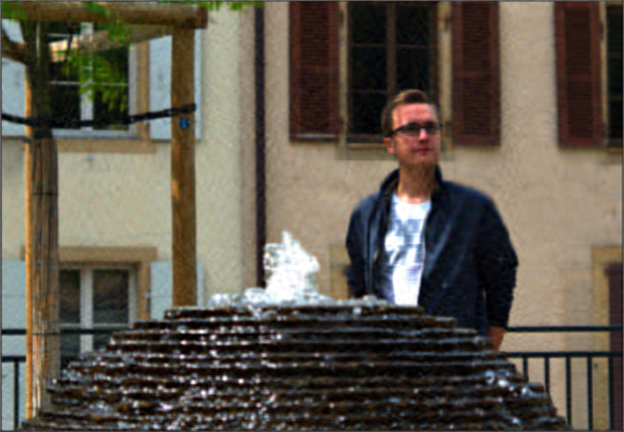}
\end{subfigure}

\begin{subfigure}{0.13\textwidth}
  \includegraphics[width=\linewidth]{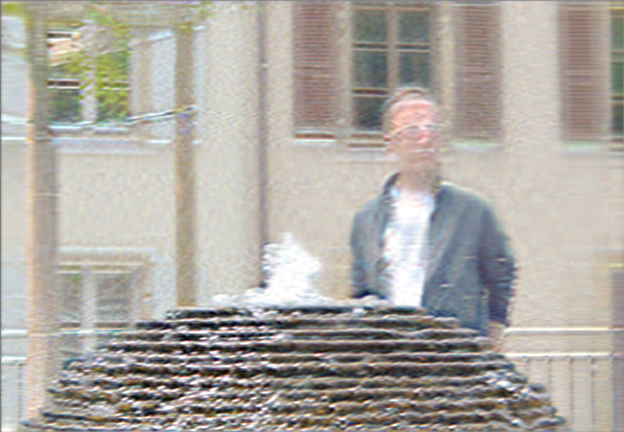}
  \caption{\footnotesize $C_{2}-S1$}
\end{subfigure} 
\begin{subfigure}{0.13\textwidth}
  \includegraphics[width=\linewidth]{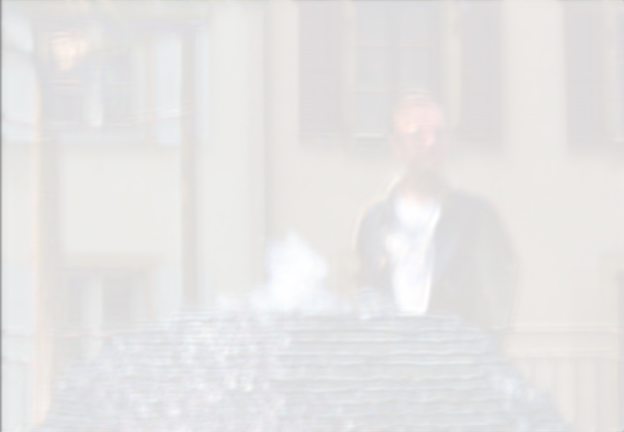}
  \caption{\footnotesize $C_{2}-S2$}
\end{subfigure} 
\begin{subfigure}{0.13\textwidth}
  \includegraphics[width=\linewidth]{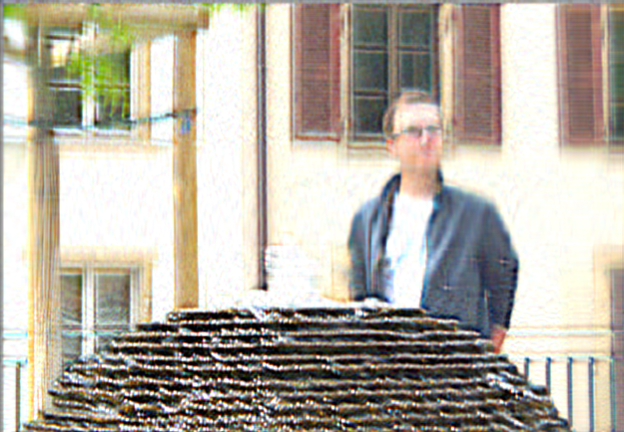}
  \caption{\footnotesize $C_{4}-S1$}
\end{subfigure}
\begin{subfigure}{0.13\textwidth}
  \includegraphics[width=\linewidth]{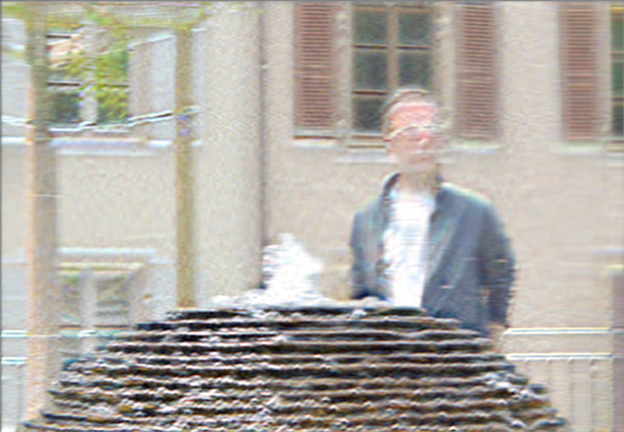}
  \caption{\footnotesize $C_{4}-S2$}
\end{subfigure} 
\begin{subfigure}{0.13\textwidth}
  \includegraphics[width=\linewidth]{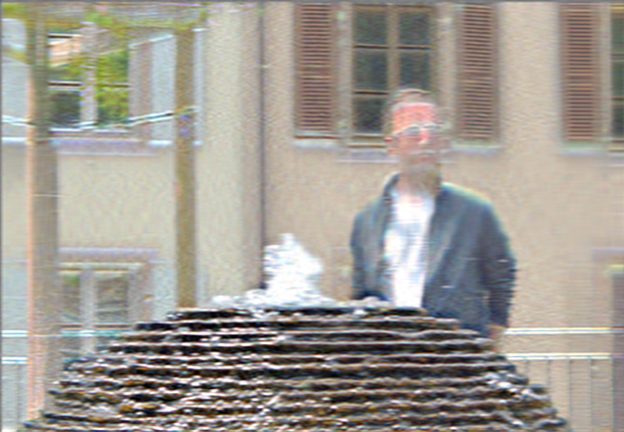}
  \caption{\footnotesize $C_{4}-S3$}
\end{subfigure} 
\begin{subfigure}{0.13\textwidth}
  \includegraphics[width=\linewidth]{./Figures/c4_s3_f02.png}
  \caption{\footnotesize $C_{4}-S4$}
\end{subfigure}

\caption{\footnotesize The multiplicative layers of each view subset (subset 1 to 4 denoted as S1 to S4) of the Circular-2 ($C_2$) and Circular-4 ($C_4)$ scanning patterns. The first, second and third rows illustrate multiplicative layers -1. 0 and 1 respectively. }
\label{fig:fountlayers}
\end{figure*}

\begin{figure*}
\centering
\begin{subfigure}{0.13\textwidth}
  \includegraphics[width=\linewidth]{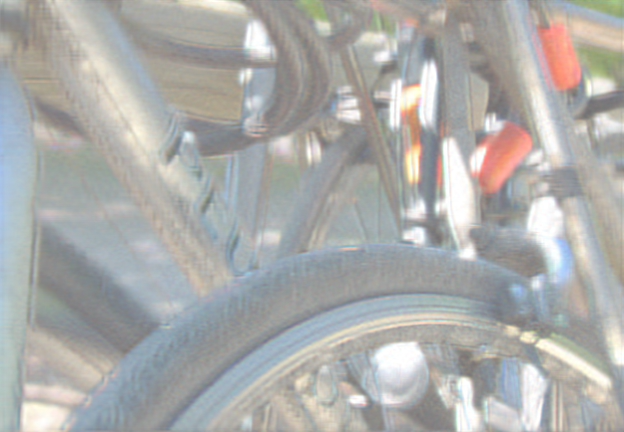}
\end{subfigure} 
\begin{subfigure}{0.13\textwidth}
  \includegraphics[width=\linewidth]{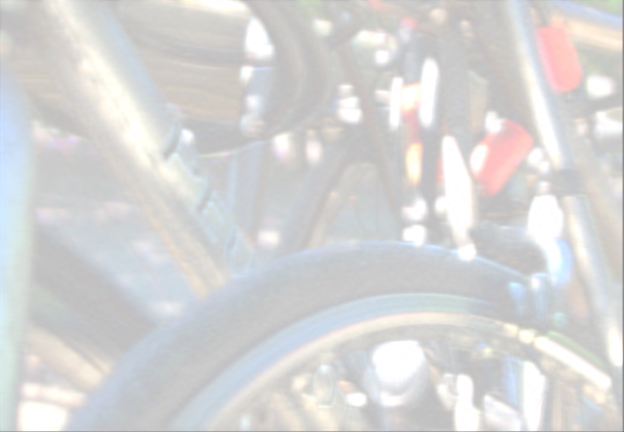}
\end{subfigure} 
\begin{subfigure}{0.13\textwidth}
  \includegraphics[width=\linewidth]{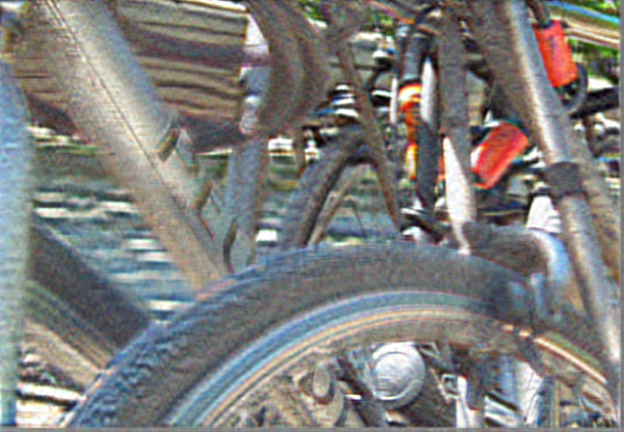}
\end{subfigure}
\begin{subfigure}{0.13\textwidth}
  \includegraphics[width=\linewidth]{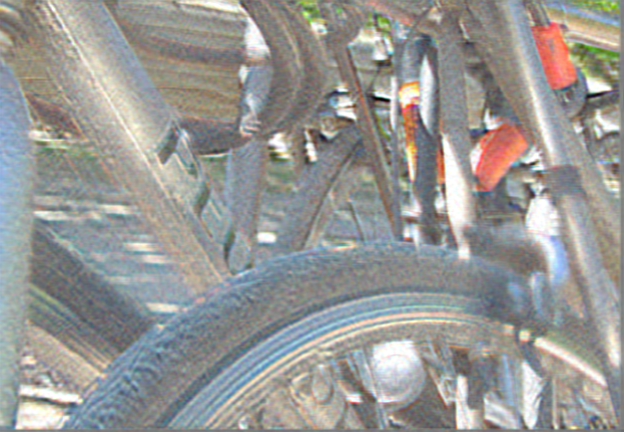}
\end{subfigure} 
\begin{subfigure}{0.13\textwidth}
  \includegraphics[width=\linewidth]{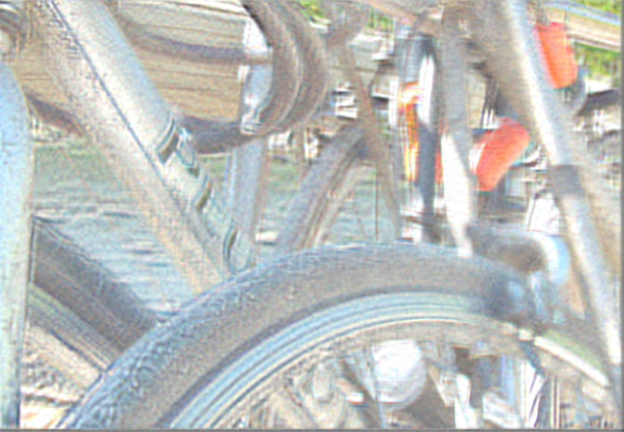}
\end{subfigure} 
\begin{subfigure}{0.13\textwidth}
  \includegraphics[width=\linewidth]{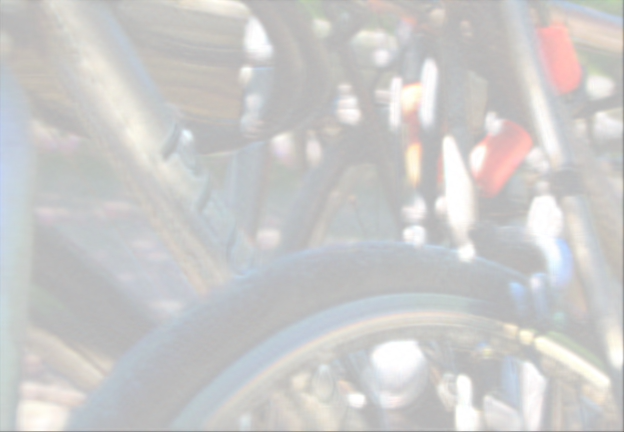}
\end{subfigure}

\begin{subfigure}{0.13\textwidth}
  \includegraphics[width=\linewidth]{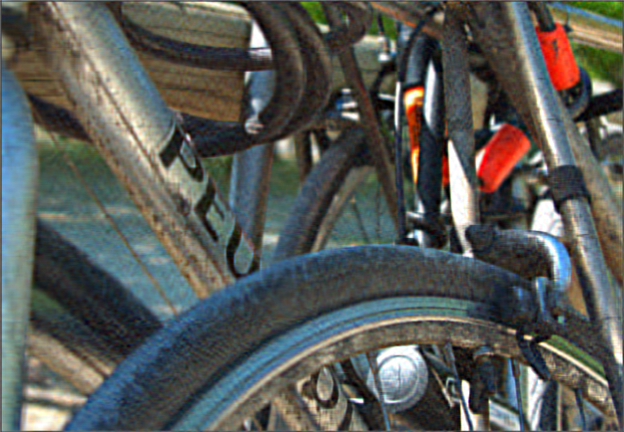}
\end{subfigure} 
\begin{subfigure}{0.13\textwidth}
  \includegraphics[width=\linewidth]{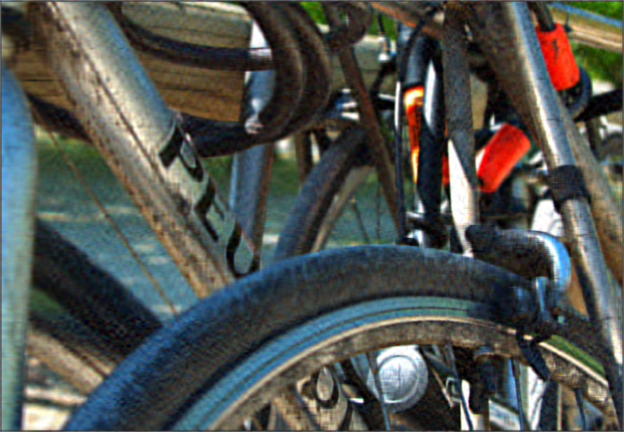}
\end{subfigure} 
\begin{subfigure}{0.13\textwidth}
  \includegraphics[width=\linewidth]{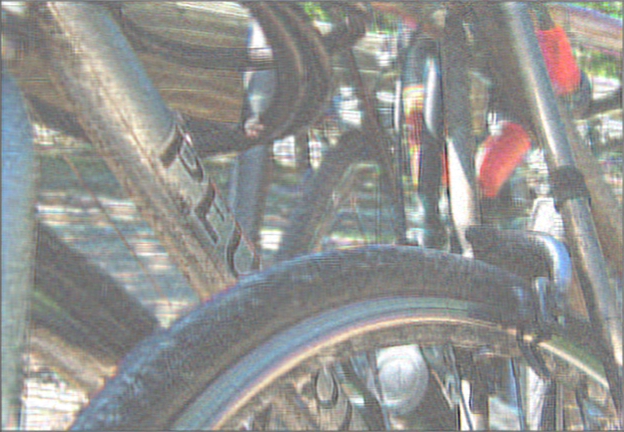}
\end{subfigure}
\begin{subfigure}{0.13\textwidth}
  \includegraphics[width=\linewidth]{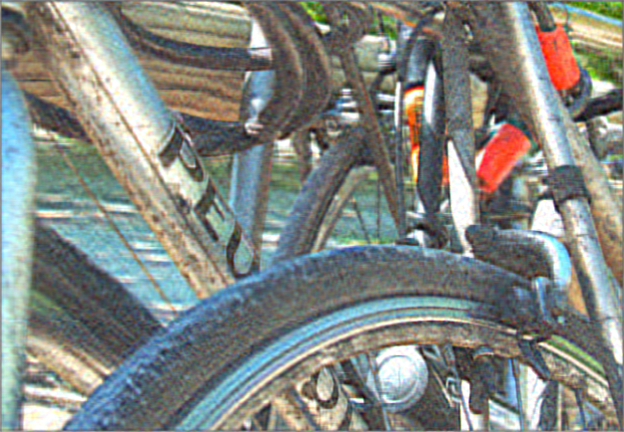}
\end{subfigure} 
\begin{subfigure}{0.13\textwidth}
  \includegraphics[width=\linewidth]{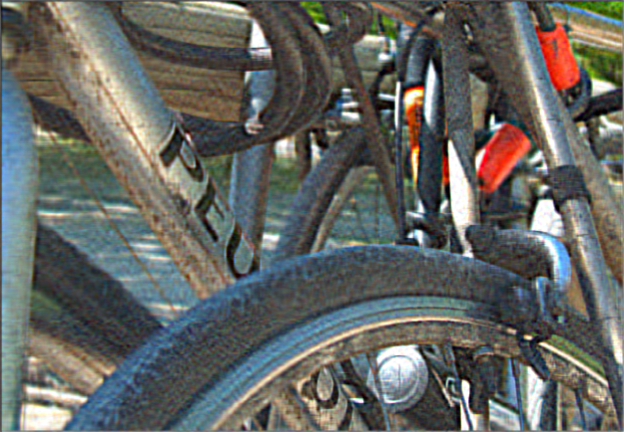}
\end{subfigure} 
\begin{subfigure}{0.13\textwidth}
  \includegraphics[width=\linewidth]{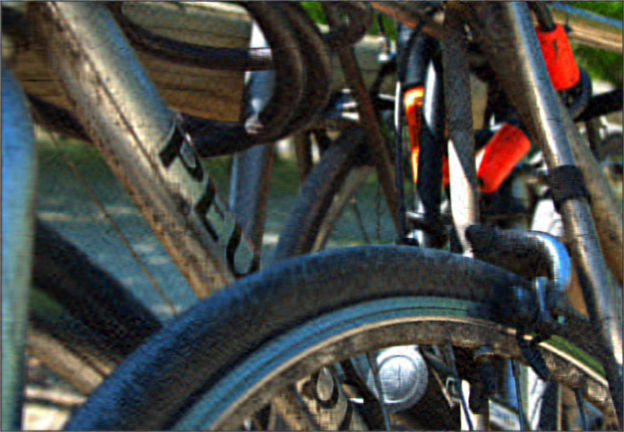}
\end{subfigure}

\begin{subfigure}{0.13\textwidth}
  \includegraphics[width=\linewidth]{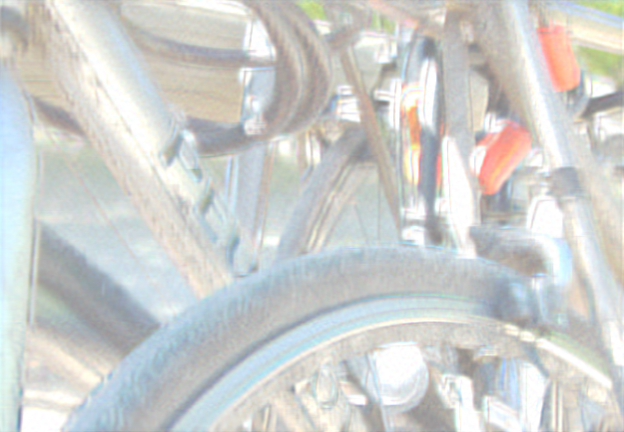}
  \caption{\footnotesize $H_{2}-S1$}
\end{subfigure} 
\begin{subfigure}{0.13\textwidth}
  \includegraphics[width=\linewidth]{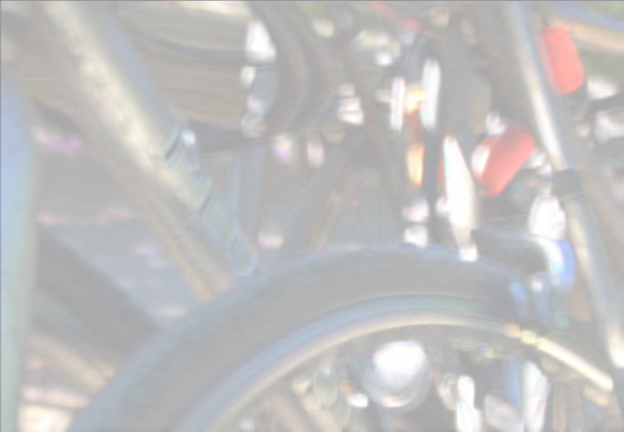}
  \caption{\footnotesize $H_{2}-S2$}
\end{subfigure} 
\begin{subfigure}{0.13\textwidth}
  \includegraphics[width=\linewidth]{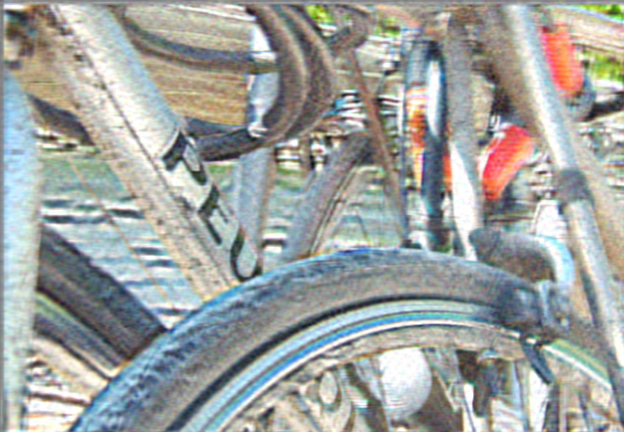}
  \caption{\footnotesize $H_{4}-S1$}
\end{subfigure}
\begin{subfigure}{0.13\textwidth}
  \includegraphics[width=\linewidth]{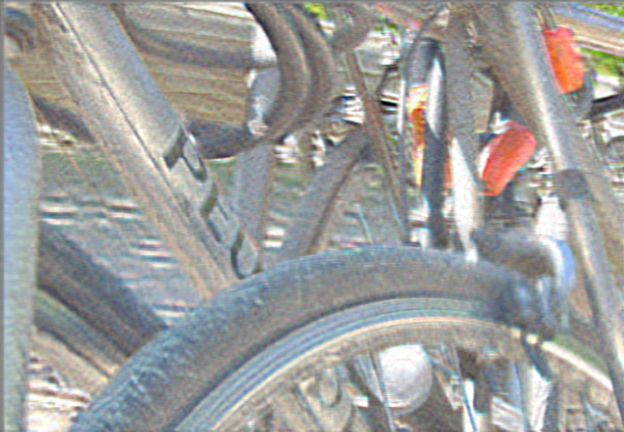}
  \caption{\footnotesize $H_{4}-S2$}
\end{subfigure} 
\begin{subfigure}{0.13\textwidth}
  \includegraphics[width=\linewidth]{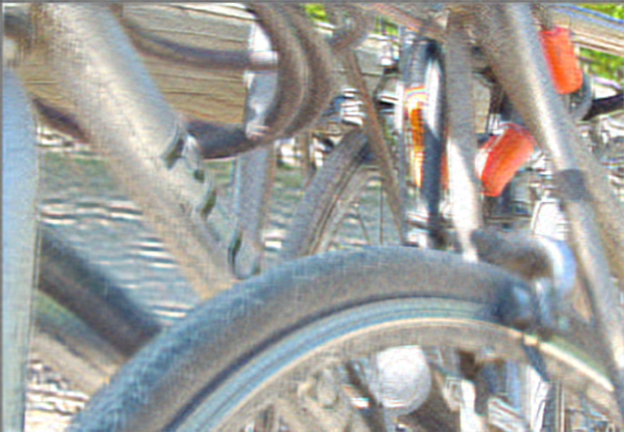}
  \caption{\footnotesize $H_{4}-S3$}
\end{subfigure} 
\begin{subfigure}{0.13\textwidth}
  \includegraphics[width=\linewidth]{./Figures/h4_s3_b02.png}
  \caption{\footnotesize $H_{4}-S4$}
\end{subfigure}

\caption{\footnotesize The multiplicative layers of each view subset (subset 1 to 4 denoted as S1 to S4) of the Hierarchical-2 ($H_2$) and Hierarchical-4 ($H_4)$ scanning patterns. The first, second and third rows illustrate multiplicative layers -1. 0 and 1 respectively. }
\label{fig:bikeslayers}
\end{figure*}


\begin{figure*}
\centering
\begin{subfigure}{0.32\linewidth}
  \includegraphics[width=\linewidth]{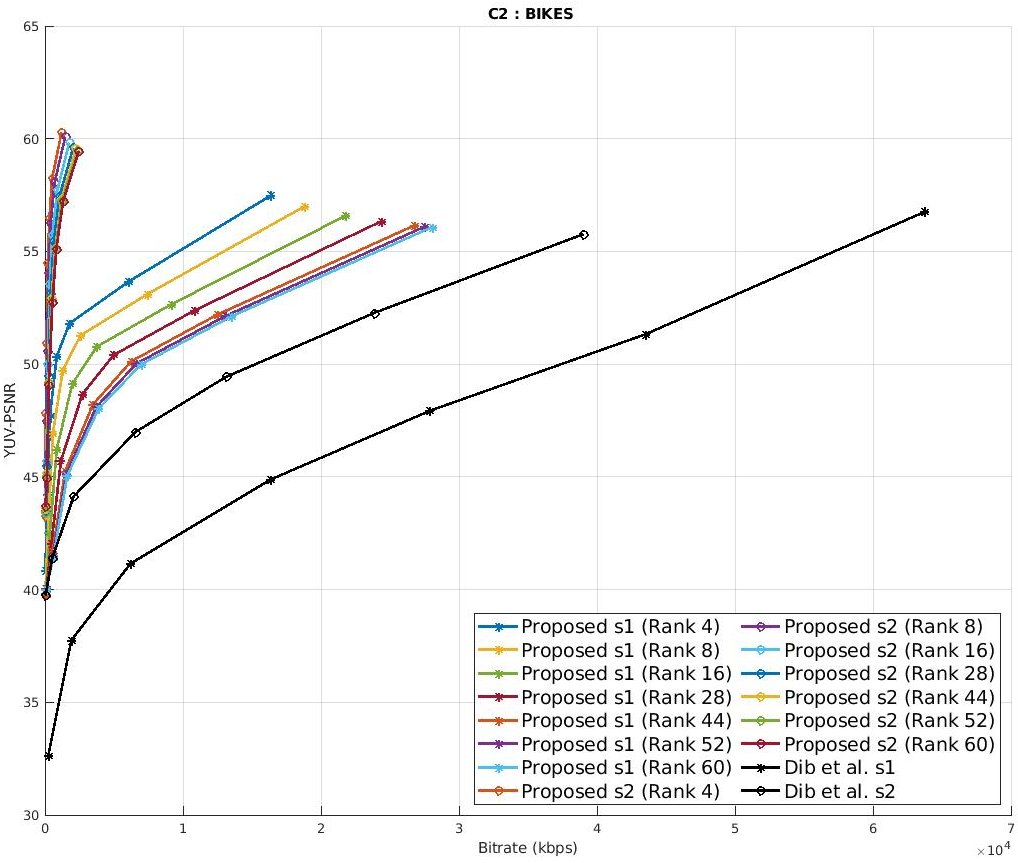}
\end{subfigure} 
\begin{subfigure}{0.32\linewidth}
  \includegraphics[width=\linewidth]{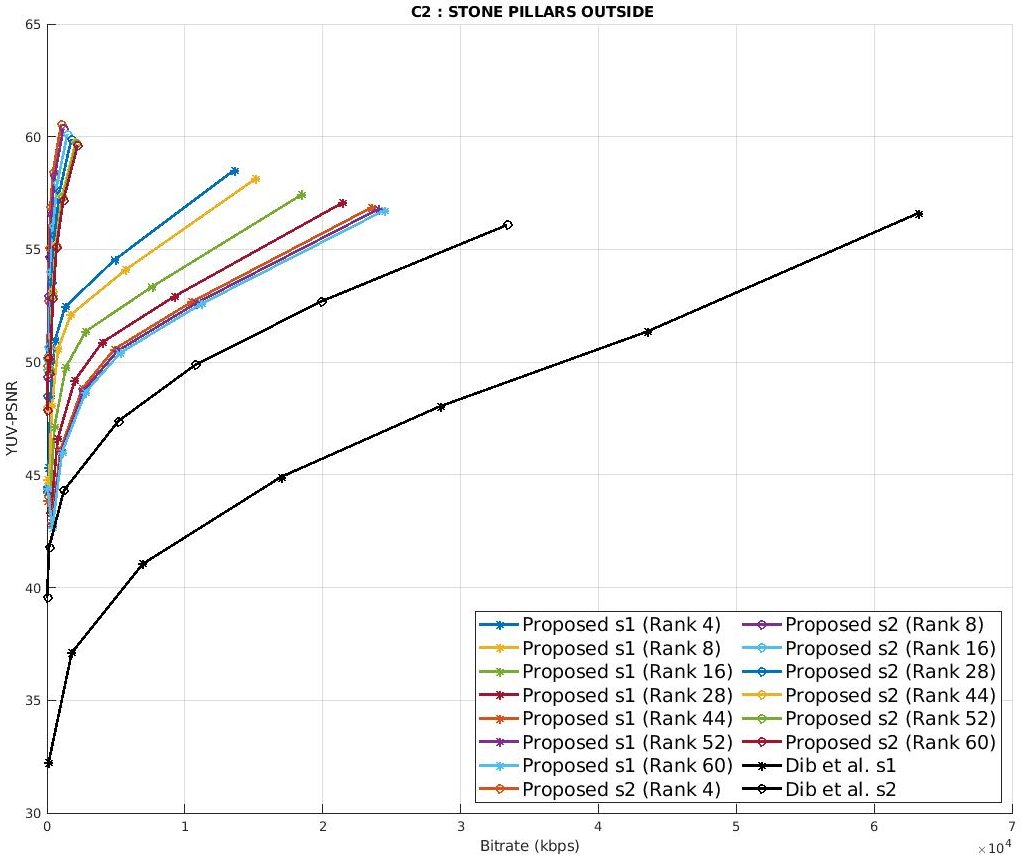}
\end{subfigure}
\begin{subfigure}{0.32\linewidth}
  \includegraphics[width=\linewidth]{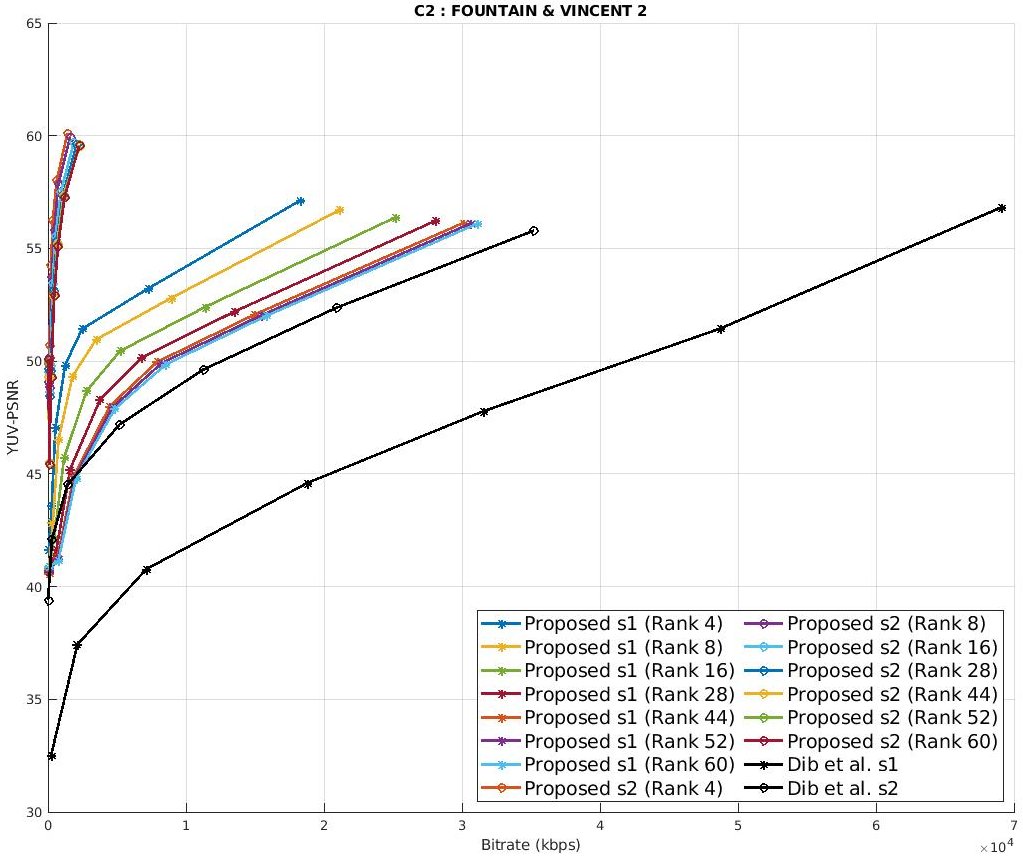}
\end{subfigure}

\begin{subfigure}{0.32\linewidth}
  \includegraphics[width=\linewidth]{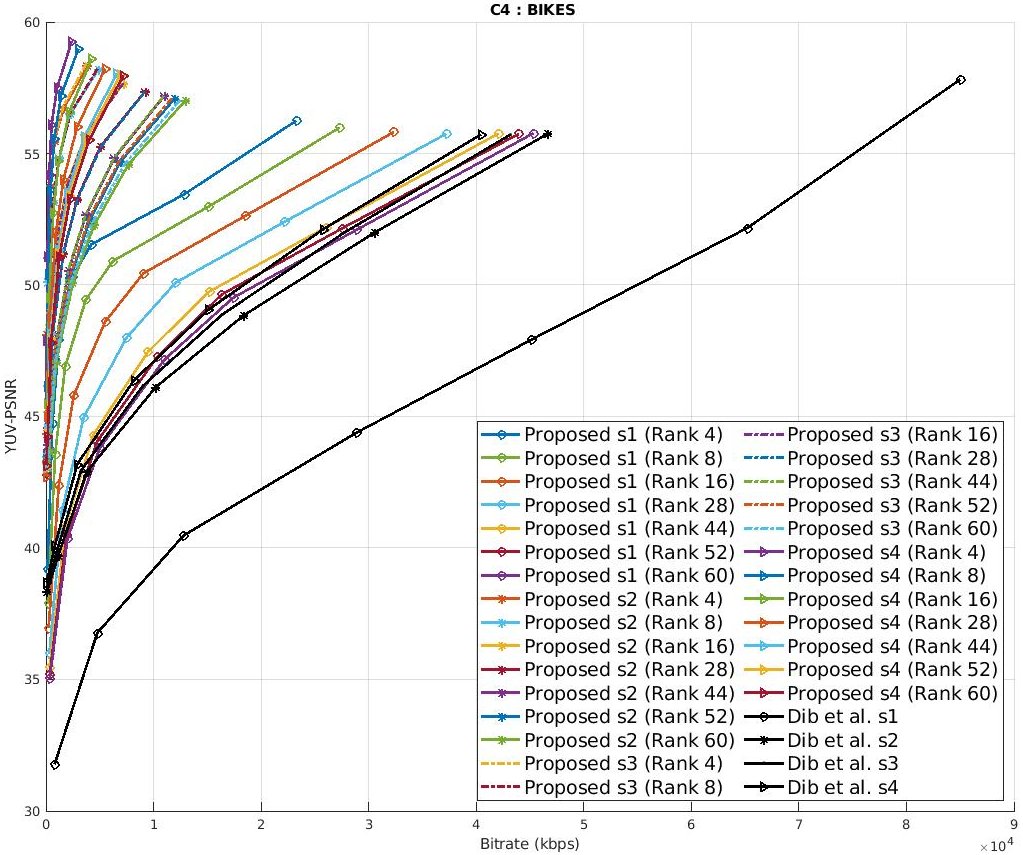}
\end{subfigure} 
\begin{subfigure}{0.32\linewidth}
  \includegraphics[width=\linewidth]{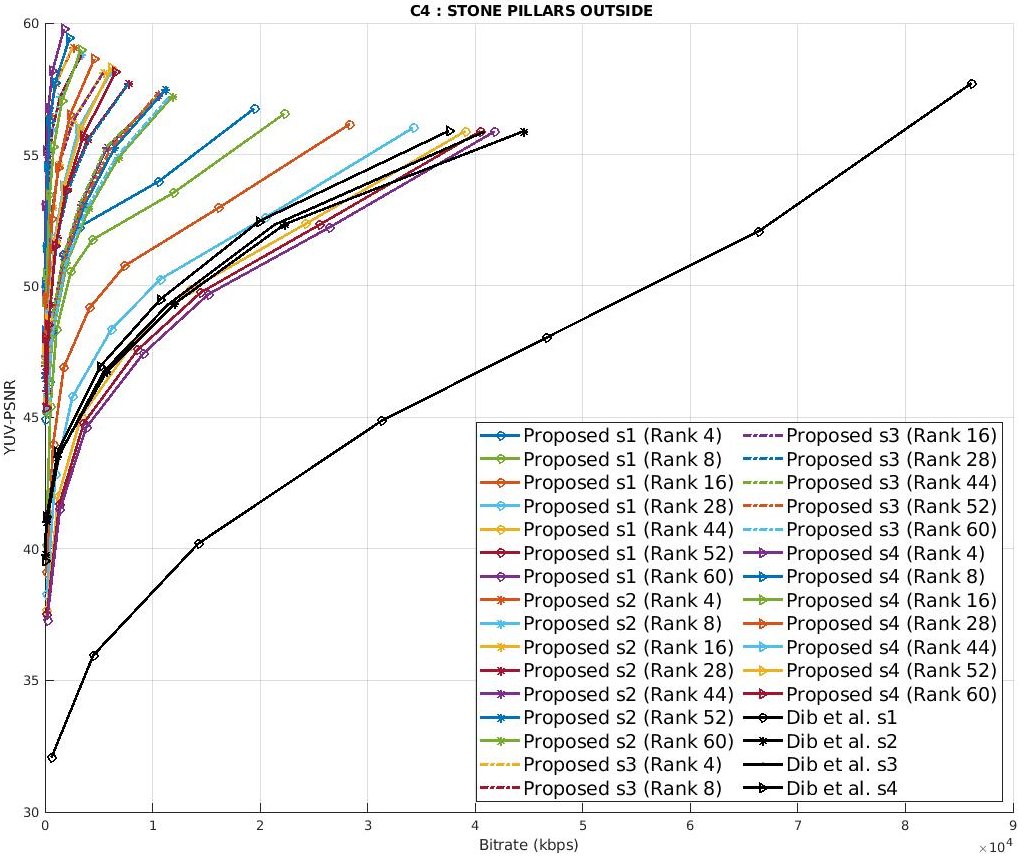}
\end{subfigure}
\begin{subfigure}{0.32\linewidth}
  \includegraphics[width=\linewidth]{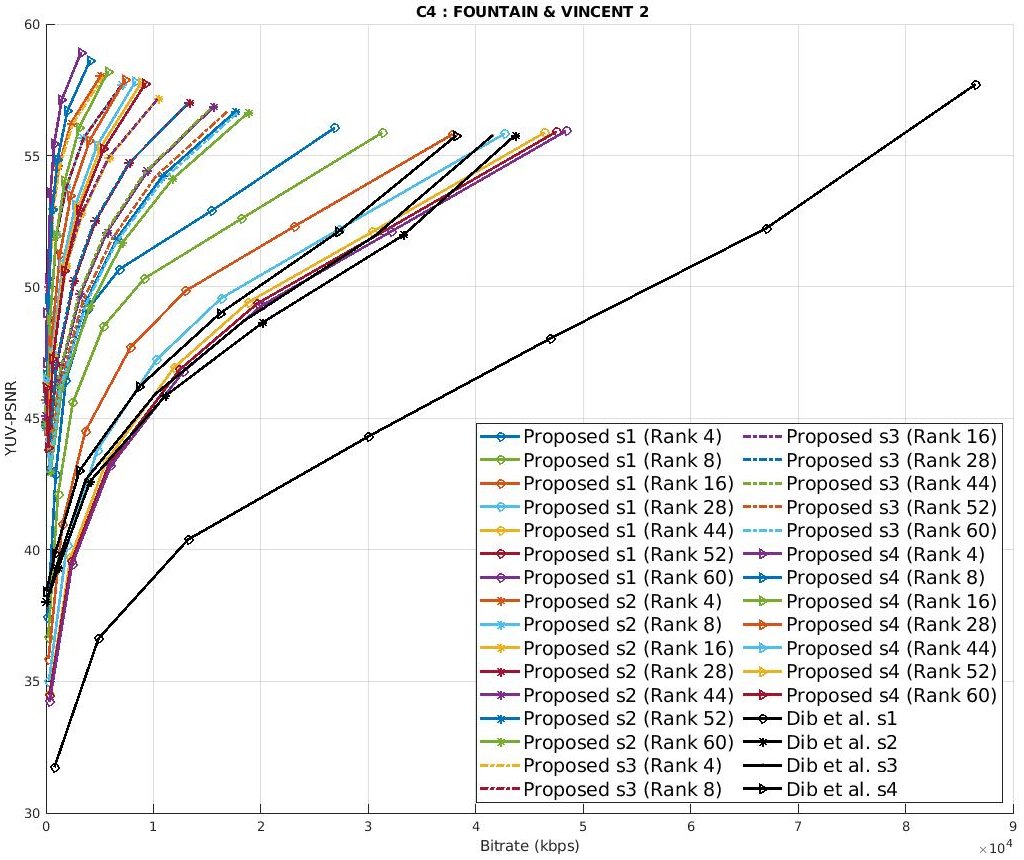}
\end{subfigure}

\begin{subfigure}{0.32\linewidth}
  \includegraphics[width=\linewidth]{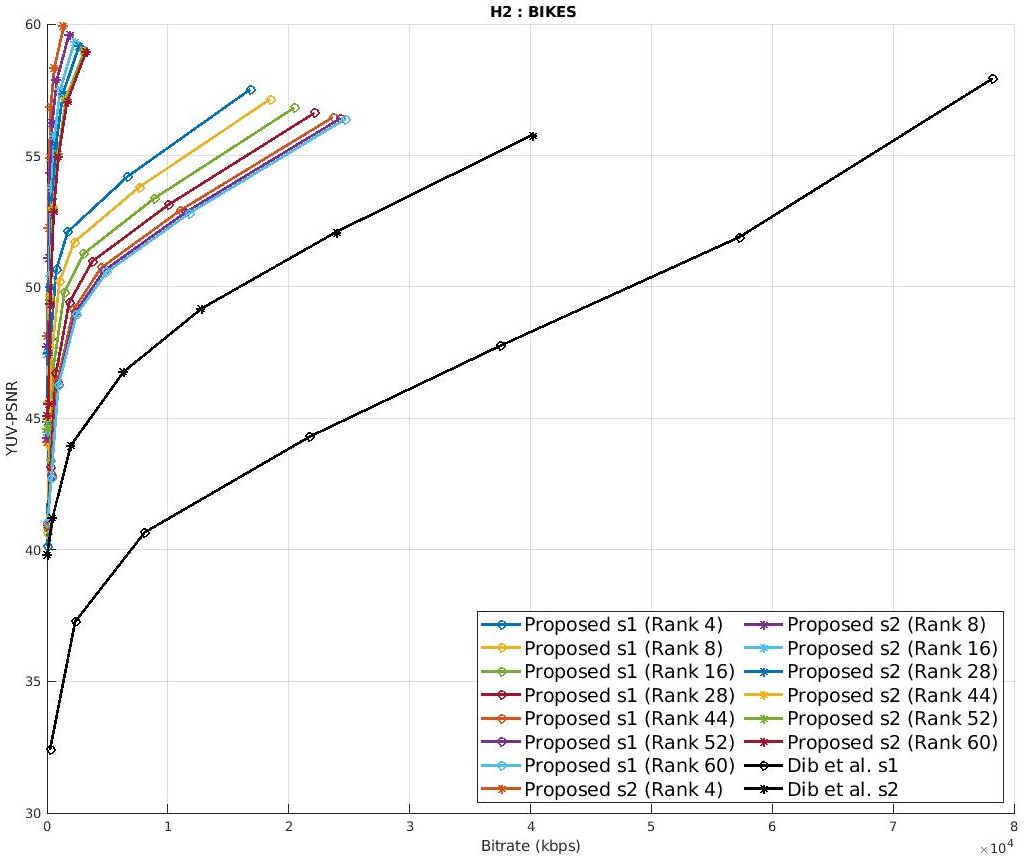}
\end{subfigure} 
\begin{subfigure}{0.32\linewidth}
  \includegraphics[width=\linewidth]{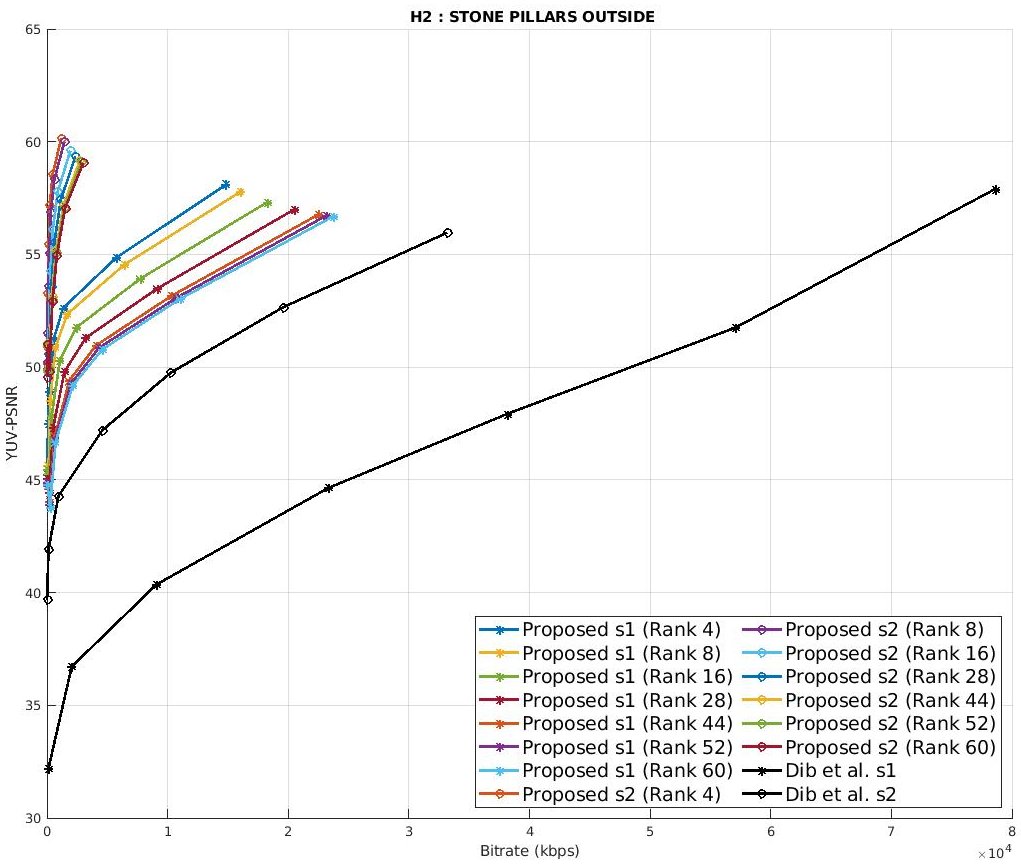}
\end{subfigure}
\begin{subfigure}{0.32\linewidth}
  \includegraphics[width=\linewidth]{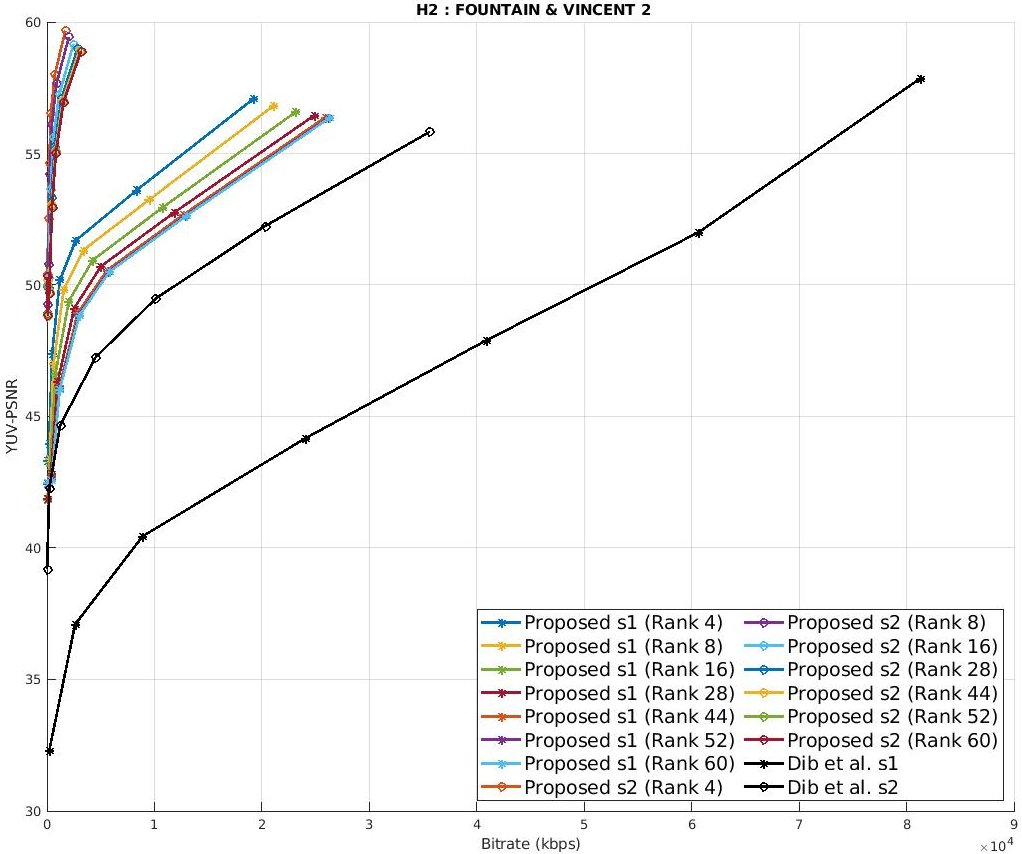}
\end{subfigure} 

\begin{subfigure}{0.32\linewidth}
  \includegraphics[width=\linewidth]{./Figures/h2_bikes.jpg}
    \caption{\footnotesize Bikes}
\end{subfigure} 
\begin{subfigure}{0.32\linewidth}
  \includegraphics[width=\linewidth]{./Figures/h2_stone.jpg}
  \caption{\footnotesize Stone pillars Outside}
\end{subfigure}
\begin{subfigure}{0.32\linewidth}
  \includegraphics[width=\linewidth]{./Figures/h2_fount.jpg}
  \caption{\footnotesize  Fountain \& Vincent 2}
\end{subfigure} 

\caption{\footnotesize Bitrate vs YUV-PSNR curves for the proposed compression scheme and Dib et al. for the three datasets in the $C_2$, $C_4$, $H_2$ and $H_4$ patterns. \textit{(Kindly expand for better clarity)}}
\label{fig:graphs}
\end{figure*}

\subsubsection{FDL View Synthesis and Prediction}

In section~\ref{sec:view_subsets}, two basic scanning orders, circular and hierarchical, are discussed for the synthesis and coding of light field views. Each of these chosen orders has two or four view subsets. This results in four patterns $C_2$, $C_4$, $H_2$ and $H_4$. In all four cases of the view prediction orders, the images are arranged in a spiral order starting from the center of the light field for each subset. The initial view subset of every pattern is always the first subset as specified by the scanning order. This subset is directly encoded first in COMPONENT II. For example, in Fig.~\ref{fig:view subset orders}, the blue coloured subset in $C_2$ is the first subset.

The angular coordinates $u_j$ and disparity values $d_k$ are determined by the Fourier Disparity Layer calibration~\cite{le2019fourier}. These are required in the further FDL construction and view predictions. This additional information is transmitted to the decoder in COMPONENT III as metadata~\cite{dib2019light}. The initial view subset is used in the basic construction of FDL representation. This aids in the synthesis of succeeding view subsets. The residual signal is also encoded to account for the remaining correlations in the prediction residue of synthesized views. The FDL representation is then refined before prediction and encoding of the next subset of views. Thus, the FDL representation is iteratively fine-tuned at every stage, after encoding every view subset, until all the approximated input light field views are encoded.


\section{Results and Analysis}\label{ra}
The performance of the proposed compression scheme is evaluated on real light fields captured by plenoptic cameras. The experiments are performed with  \textit{Bikes}, \textit{Fountain \& Vincent 2}, and \textit{Stone pillars outside} light field datasets from the EPFL Lightfield JPEG Pleno database~\cite{rerabek2016new}. The central views of the chosen light field images are shown in Fig.~\ref{fig:orglfs}. The raw plenoptic images are extracted into 15$\times$15 sub-aperture views using the Matlab Light field toolbox~\cite{dansereau2013decoding}.

The patterns $C_2$, $C_4$, $H_2$ and $H_4$ are constructed from inner 9$\times$9 light field views for our experiments. Subsets 1 and 2 of $C_2$ contain 24 and 57 light field views respectively. The first and second subsets $H_2$ contain 25 and 56 views respectively. In $C_4$, subsets 1, 2, 3 and 4 have 4, 16, 12 and 49 views respectively. Lastly, subsets 1, 2, 3 and 4 of $H_4$ contain 4, 5, 16 and 56 light field views respectively. The exact scanning orders of the patterns and their subsets are specified in Fig.~\ref{fig:view subset orders}.

\begin{table}
\centering
\caption{\footnotesize Total number of bytes for each subset of Circular-2 pattern using our proposed coding scheme and Dib et al. \cite{dib2019light}.}
\label{table_bytesC2}
\resizebox{6cm}{!}{
\makebox[\linewidth]{\begin{tabular}{|c|c|c|c|c|c|c|c|}
\hline
\multicolumn{2}{|c|}{} & \multicolumn{2}{|c|}{Bikes} & \multicolumn{2}{|c|}{Stone pillars outside}  & \multicolumn{2}{|c|}{Fountain \& Vincent 2} \\
\hline
QP & Scheme & Subset 1 & Subset 2 & Subset 1 & Subset 2 & Subset 1 & Subset 2\\
\hline
\multirow{8}{*}{2} & Dib et al.	&	6373276	&	9269248	&	6321523	&	7928095	&	6912472	&	8363327	\\	\cline{2-8}
&	Proposed (Rank 4)	&	1634865	&	285016	&	1358448	&	243122	&	1827059	&	335430	\\	\cline{2-8}
&	Proposed (Rank 8)	&	1877550	&	349216	&	1512085	&	283331	&	2115774	&	380847	\\	\cline{2-8}
&	Proposed (Rank 16)	&	2177689	&	417606	&	1847285	&	351680	&	2519102	&	446949	\\	\cline{2-8}
&	Proposed (Rank 28)	&	2438540	&	486569	&	2142502	&	418877	&	2806767	&	494263	\\	\cline{2-8}
&	Proposed (Rank 44)	&	2674089	&	532163	&	2350849	&	487484	&	3008060	&	545947	\\	\cline{2-8}
&	Proposed (Rank 52)	&	2749844	&	561758	&	2401859	&	503039	&	3063768	&	535674	\\	\cline{2-8}
&	Proposed (Rank 60)	&	2805401	&	581882	&	2448047	&	527789	&	3113991	&	535619	\\	\cline{1-8}
\hline
\multirow{8}{*}{6} &	Dib et al.	&	4350810	&	5663692	&	4354754	&	4732850	&	4869817	&	4963030	\\	\cline{2-8}
&	Proposed (Rank 4)	&	602334	&	124074	&	493765	&	104272	&	728339	&	149454	\\	\cline{2-8}
&	Proposed (Rank 8)	&	741293	&	165605	&	571942	&	130267	&	893240	&	178163	\\	\cline{2-8}
&	Proposed (Rank 16)	&	916846	&	211107	&	761895	&	173837	&	1138839	&	220365	\\	\cline{2-8}
&	Proposed (Rank 28)	&	1086119	&	247005	&	925727	&	218830	&	1349095	&	252514	\\	\cline{2-8}
&	Proposed (Rank 44)	&	1252891	&	282873	&	1051703	&	256343	&	1499318	&	268165	\\	\cline{2-8}
&	Proposed (Rank 52)	&	1311143	&	290011	&	1084796	&	268153	&	1547468	&	279455	\\	\cline{2-8}
&	Proposed (Rank 60)	&	1354933	&	310371	&	1118995	&	281491	&	1583718	&	284981	\\	\cline{1-8}
\hline
\multirow{8}{*}{10} &	Dib et al.	&	4350810	&	5663692	&	4354754	&	4732850	&	4869817	&	4963030	\\	\cline{2-8}
&	Proposed (Rank 4)	&	602334	&	124074	&	493765	&	104272	&	728339	&	149454	\\	\cline{2-8}
&	Proposed (Rank 8)	&	741293	&	165605	&	571942	&	130267	&	893240	&	178163	\\	\cline{2-8}
&	Proposed (Rank 16)	&	916846	&	211107	&	761895	&	173837	&	1138839	&	220365	\\	\cline{2-8}
&	Proposed (Rank 28)	&	1086119	&	247005	&	925727	&	218830	&	1349095	&	252514	\\	\cline{2-8}
&	Proposed (Rank 44)	&	1252891	&	282873	&	1051703	&	256343	&	1499318	&	268165	\\	\cline{2-8}
&	Proposed (Rank 52)	&	1311143	&	290011	&	1084796	&	268153	&	1547468	&	279455	\\	\cline{2-8}
&	Proposed (Rank 60)	&	1354933	&	310371	&	1118995	&	281491	&	1583718	&	284981	\\	\cline{1-8}
\hline
\multirow{8}{*}{14} &	Dib et al.	&	1635624	&	1555027	&	1697516	&	1227782	&	1879832	&	1231959	\\	\cline{2-8}
&	Proposed (Rank 4)	&	85418	&	47842	&	60119	&	36583	&	124338	&	51127	\\	\cline{2-8}
&	Proposed (Rank 8)	&	129326	&	64433	&	81540	&	45979	&	177036	&	66168	\\	\cline{2-8}
&	Proposed (Rank 16)	&	197901	&	80959	&	136933	&	64691	&	279887	&	79125	\\	\cline{2-8}
&	Proposed (Rank 28)	&	271565	&	98486	&	200957	&	79562	&	376807	&	94429	\\	\cline{2-8}
&	Proposed (Rank 44)	&	347606	&	112747	&	255491	&	93448	&	449040	&	103222	\\	\cline{2-8}
&	Proposed (Rank 52)	&	372263	&	119795	&	267999	&	97639	&	466826	&	103998	\\	\cline{2-8}
&	Proposed (Rank 60)	&	389902	&	127835	&	280325	&	102454	&	483184	&	107398	\\	\cline{1-8}
\hline
\multirow{8}{*}{20} &	Dib et al.	&	1635624	&	1555027	&	695581	&	292728	&	713369	&	335902	\\	\cline{2-8}
&	Proposed (Rank 4)	&	85418	&	47842	&	25871	&	19602	&	54329	&	25826	\\	\cline{2-8}
&	Proposed (Rank 8)	&	129326	&	64433	&	34316	&	23444	&	76408	&	33557	\\	\cline{2-8}
&	Proposed (Rank 16)	&	197901	&	80959	&	55115	&	31584	&	116924	&	40566	\\	\cline{2-8}
&	Proposed (Rank 28)	&	271565	&	98486	&	77665	&	38192	&	160911	&	47636	\\	\cline{2-8}
&	Proposed (Rank 44)	&	347606	&	112747	&	99503	&	44221	&	187776	&	51037	\\	\cline{2-8}
&	Proposed (Rank 52)	&	372263	&	119795	&	104642	&	46549	&	195728	&	52069	\\	\cline{2-8}
&	Proposed (Rank 60)	&	389902	&	127835	&	109416	&	48116	&	203384	&	51973	\\	\cline{1-8}
\hline
\multirow{8}{*}{26} &	Dib et al.	&	191049	&	125748	&	180485	&	36164	&	213553	&	66399	\\	\cline{2-8}
&	Proposed (Rank 4)	&	18109	&	15295	&	12194	&	10550	&	26763	&	14884	\\	\cline{2-8}
&	Proposed (Rank 8)	&	26471	&	20342	&	16372	&	13010	&	35126	&	18824	\\	\cline{2-8}
&	Proposed (Rank 16)	&	36035	&	24566	&	22220	&	15525	&	46554	&	19651	\\	\cline{2-8}
&	Proposed (Rank 28)	&	46328	&	28090	&	28818	&	17922	&	60959	&	23291	\\	\cline{2-8}
&	Proposed (Rank 44)	&	55291	&	29746	&	35061	&	20507	&	70030	&	24332	\\	\cline{2-8}
&	Proposed (Rank 52)	&	58525	&	30969	&	37056	&	20533	&	73670	&	24606	\\	\cline{2-8}
&	Proposed (Rank 60)	&	61024	&	33335	&	39291	&	21200	&	75358	&	25193	\\	\cline{1-8}
\hline
\multirow{8}{*}{38} &	Dib et al.	&	23065	&	7880	&	12613	&	1880	&	22582	&	2155	\\	\cline{2-8}
&	Proposed (Rank 4)	&	2737	&	3283	&	2289	&	3043	&	5098	&	3887	\\	\cline{2-8}
&	Proposed (Rank 8)	&	4208	&	4244	&	2719	&	3314	&	6606	&	4240	\\	\cline{2-8}
&	Proposed (Rank 16)	&	5752	&	5095	&	3197	&	3572	&	7206	&	4708	\\	\cline{2-8}
&	Proposed (Rank 28)	&	6864	&	5757	&	3888	&	3914	&	8261	&	4785	\\	\cline{2-8}
&	Proposed (Rank 44)	&	7852	&	6246	&	4442	&	3969	&	9137	&	5104	\\	\cline{2-8}
&	Proposed (Rank 52)	&	7946	&	6244	&	4437	&	4024	&	9294	&	5126	\\	\cline{2-8}
&	Proposed (Rank 60)	&	8559	&	6592	&	4553	&	4038	&	9630	&	5197	\\	\cline{1-8}
\hline
\end{tabular} }
}
\end{table}

\begin{table}
\centering
\caption{\footnotesize Total number of bytes for each subset of Hierarchical-2 pattern using our proposed coding scheme and Dib et al. \cite{dib2019light}. }
\label{table_bytesH2}
\resizebox{6cm}{!}{
\makebox[\linewidth]{\begin{tabular}{|c|c|c|c|c|c|c|c|}
\hline
\multicolumn{2}{|c|}{} & \multicolumn{2}{|c|}{Bikes} & \multicolumn{2}{|c|}{Stone pillars outside}  & \multicolumn{2}{|c|}{Fountain \& Vincent 2} \\
\hline
QP & Scheme & Subset 1 & Subset 2 & Subset 1 & Subset 2 & Subset 1 & Subset 2\\
\hline
\multirow{8}{*}{2} &	Dib et al.	&	8149470	&	9362963	&	8187441	&	7746889	&	8467600	&	8311549	\\	\cline{2-8}
&	Proposed (Rank 4)	&	1752364	&	327571	&	1544489	&	277441	&	2000768	&	399280	\\	\cline{2-8}
&	Proposed (Rank 8)	&	1923491	&	425176	&	1668430	&	336414	&	2197190	&	475908	\\	\cline{2-8}
&	Proposed (Rank 16)	&	2132305	&	534173	&	1904622	&	452160	&	2413070	&	576114	\\	\cline{2-8}
&	Proposed (Rank 28)	&	2307124	&	626191	&	2137972	&	550384	&	2589097	&	663419	\\	\cline{2-8}
&	Proposed (Rank 44)	&	2466765	&	728164	&	2348147	&	638116	&	2701529	&	709257	\\	\cline{2-8}
&	Proposed (Rank 52)	&	2531862	&	732311	&	2415703	&	670368	&	2728626	&	732779	\\	\cline{2-8}
&	Proposed (Rank 60)	&	2572006	&	759606	&	2470431	&	700779	&	2741756	&	740021	\\	\cline{1-8}
\hline
\multirow{8}{*}{6} &	Dib et al.	&	5968201	&	5584732	&	5943824	&	4566311	&	6317422	&	4734277	\\	\cline{2-8}
&	Proposed (Rank 4)	&	696307	&	141510	&	600337	&	110556	&	870560	&	169079	\\	\cline{2-8}
&	Proposed (Rank 8)	&	796855	&	191515	&	665513	&	143058	&	991937	&	208585	\\	\cline{2-8}
&	Proposed (Rank 16)	&	928516	&	259814	&	809014	&	209917	&	1119092	&	266975	\\	\cline{2-8}
&	Proposed (Rank 28)	&	1044367	&	309978	&	953162	&	267569	&	1237456	&	314759	\\	\cline{2-8}
&	Proposed (Rank 44)	&	1151050	&	364678	&	1079187	&	314645	&	1320242	&	341091	\\	\cline{2-8}
&	Proposed (Rank 52)	&	1195073	&	376134	&	1122665	&	333719	&	1345621	&	355264	\\	\cline{2-8}
&	Proposed (Rank 60)	&	1224508	&	388297	&	1153867	&	350318	&	1356529	&	359455	\\	\cline{1-8}
\hline
\multirow{8}{*}{10} &	Dib et al.	&	3906437	&	2991907	&	3972408	&	2394424	&	4256777	&	2365804	\\	\cline{2-8}
&	Proposed (Rank 4)	&	178206	&	74529	&	139283	&	54783	&	278393	&	83747	\\	\cline{2-8}
&	Proposed (Rank 8)	&	234038	&	103767	&	176282	&	74433	&	350782	&	106751	\\	\cline{2-8}
&	Proposed (Rank 16)	&	313521	&	143968	&	251942	&	112280	&	442029	&	138069	\\	\cline{2-8}
&	Proposed (Rank 28)	&	389826	&	175682	&	336035	&	146248	&	523367	&	166752	\\	\cline{2-8}
&	Proposed (Rank 44)	&	465603	&	213126	&	423251	&	172780	&	582988	&	184091	\\	\cline{2-8}
&	Proposed (Rank 52)	&	493156	&	213184	&	449828	&	182623	&	600907	&	189758	\\	\cline{2-8}
&	Proposed (Rank 60)	&	515966	&	221252	&	476680	&	192152	&	606755	&	192677	\\	\cline{1-8}
\hline
\multirow{8}{*}{14} &	Dib et al.	&	2265550	&	1479998	&	2432694	&	1068158	&	2505792	&	1058776	\\	\cline{2-8}
&	Proposed (Rank 4)	&	80756	&	43910	&	56613	&	32851	&	125237	&	48892	\\	\cline{2-8}
&	Proposed (Rank 8)	&	108524	&	61888	&	74445	&	41825	&	162045	&	61496	\\	\cline{2-8}
&	Proposed (Rank 16)	&	150025	&	85871	&	111220	&	63932	&	212590	&	78435	\\	\cline{2-8}
&	Proposed (Rank 28)	&	191702	&	105961	&	151212	&	82963	&	264760	&	93719	\\	\cline{2-8}
&	Proposed (Rank 44)	&	228876	&	121783	&	194620	&	96725	&	297942	&	104203	\\	\cline{2-8}
&	Proposed (Rank 52)	&	243903	&	126699	&	209660	&	101918	&	309631	&	106855	\\	\cline{2-8}
&	Proposed (Rank 60)	&	254844	&	129017	&	222970	&	106454	&	314274	&	109546	\\	\cline{1-8}
\hline
\multirow{8}{*}{20} &	Dib et al.	&	842135	&	459805	&	946011	&	221956	&	926805	&	286174	\\	\cline{2-8}
&	Proposed (Rank 4)	&	31602	&	20923	&	21663	&	15803	&	50142	&	23547	\\	\cline{2-8}
&	Proposed (Rank 8)	&	43511	&	29373	&	28169	&	19788	&	63745	&	29074	\\	\cline{2-8}
&	Proposed (Rank 16)	&	60423	&	40866	&	40732	&	28521	&	83392	&	35879	\\	\cline{2-8}
&	Proposed (Rank 28)	&	76647	&	49200	&	53703	&	35594	&	103900	&	42129	\\	\cline{2-8}
&	Proposed (Rank 44)	&	89653	&	56477	&	66649	&	41210	&	118608	&	45865	\\	\cline{2-8}
&	Proposed (Rank 52)	&	94969	&	59779	&	71693	&	42630	&	121445	&	47138	\\	\cline{2-8}
&	Proposed (Rank 60)	&	98186	&	61299	&	75923	&	44694	&	124309	&	47441	\\	\cline{1-8}
\hline
\multirow{8}{*}{26} &	Dib et al.	&	242763	&	111266	&	213782	&	28940	&	272168	&	50062	\\	\cline{2-8}
&	Proposed (Rank 4)	&	15705	&	11349	&	9646	&	7203	&	20645	&	11951	\\	\cline{2-8}
&	Proposed (Rank 8)	&	20006	&	14619	&	12592	&	9004	&	27722	&	14167	\\	\cline{2-8}
&	Proposed (Rank 16)	&	26014	&	19172	&	15947	&	12784	&	32178	&	15899	\\	\cline{2-8}
&	Proposed (Rank 28)	&	30615	&	22252	&	19394	&	14768	&	39019	&	17819	\\	\cline{2-8}
&	Proposed (Rank 44)	&	34692	&	25694	&	23190	&	16965	&	42684	&	19317	\\	\cline{2-8}
&	Proposed (Rank 52)	&	36820	&	26082	&	25005	&	17802	&	44419	&	19885	\\	\cline{2-8}
&	Proposed (Rank 60)	&	37464	&	26717	&	26359	&	18185	&	45207	&	19866	\\	\cline{1-8}
\hline
\multirow{8}{*}{38} &	Dib et al.	&	26295	&	6202	&	13132	&	2001	&	22971	&	1975	\\	\cline{2-8}
&	Proposed (Rank 4)	&	2485	&	2434	&	2065	&	2188	&	4019	&	3065	\\	\cline{2-8}
&	Proposed (Rank 8)	&	3475	&	3205	&	2465	&	2484	&	4976	&	3523	\\	\cline{2-8}
&	Proposed (Rank 16)	&	4143	&	3742	&	2736	&	2707	&	5456	&	3699	\\	\cline{2-8}
&	Proposed (Rank 28)	&	4872	&	4364	&	3002	&	2868	&	6114	&	4137	\\	\cline{2-8}
&	Proposed (Rank 44)	&	5280	&	4572	&	3121	&	2984	&	6327	&	4141	\\	\cline{2-8}
&	Proposed (Rank 52)	&	5604	&	4810	&	3174	&	2980	&	6384	&	4192	\\	\cline{2-8}
&	Proposed (Rank 60)	&	5704	&	4866	&	3205	&	3033	&	6467	&	4291	\\	\cline{1-8}
\hline
\end{tabular} }
}
\end{table}

\begin{table*}
\centering
\caption{\footnotesize Total number of bytes for each subset of Circular-4 pattern using our proposed coding scheme and Dib et al. \cite{dib2019light}. }
\label{table_bytesC4}
\resizebox{6cm}{!}{
\makebox[\linewidth]{\begin{tabular}{|c|c|c|c|c|c|c|c|c|c|c|c|c|c|}
\hline
\multicolumn{2}{|c|}{} & \multicolumn{4}{|c|}{Bikes} & \multicolumn{4}{|c|}{Stone pillars outside}  & \multicolumn{4}{|c|}{Fountain \& Vincent 2} \\
\hline
QP & Scheme & Subset 1 & Subset 2 & Subset 3 & Subset 4 & Subset 1 & Subset 2 & Subset 3 & Subset 4 & Subset 1 & Subset 2 & Subset 3 & Subset 4\\
\hline
\multirow{8}{*}{2} &	Dib E et al.	&	1416866	&	3109571	&	5037933	&	12979009	&	1435611	&	2966575	&	4753968	&	12051303	&	1441765	&	2912377	&	4845501	&	12249516	\\	\cline{2-14}
&	Proposed (Rank 4)	&	388099	&	258152	&	440152	&	768518	&	323881	&	173763	&	298852	&	569007	&	448166	&	341166	&	624766	&	1047411	\\	\cline{2-14}
&	Proposed (Rank 8)	&	455053	&	324627	&	558934	&	968901	&	370490	&	225967	&	392536	&	740275	&	521873	&	472205	&	817319	&	1322771	\\	\cline{2-14}
&	Proposed (Rank 16)	&	539142	&	479586	&	839084	&	1367088	&	472612	&	361397	&	632487	&	1084026	&	630974	&	695930	&	1211480	&	1864056	\\	\cline{2-14}
&	Proposed (Rank 28)	&	620618	&	615680	&	1076100	&	1771137	&	570270	&	515012	&	906551	&	1475466	&	711522	&	887817	&	1536556	&	2377277	\\	\cline{2-14}
&	Proposed (Rank 44)	&	701312	&	733863	&	1277639	&	2111200	&	651481	&	705884	&	1222296	&	1979756	&	773943	&	1039250	&	1785773	&	2679246	\\	\cline{2-14}
&	Proposed (Rank 52)	&	732590	&	796888	&	1366241	&	2220068	&	675245	&	749842	&	1245301	&	1983765	&	792259	&	1177083	&	1968347	&	2854259	\\	\cline{2-14}
&	Proposed (Rank 60)	&	755916	&	864119	&	1454924	&	2314135	&	696777	&	789366	&	1336327	&	2114408	&	806779	&	1256248	&	2078412	&	2981898	\\	\cline{1-14}
\hline
\multirow{8}{*}{6} &	Dib E et al.	&	1086423	&	2038371	&	3237999	&	8276077	&	1106257	&	1486520	&	2490436	&	6394743	&	1117362	&	2219082	&	3574509	&	8740608	\\	\cline{2-14}
&	Proposed (Rank 4)	&	214122	&	110706	&	188813	&	331799	&	175603	&	67554	&	116720	&	226090	&	256811	&	156184	&	289118	&	471852	\\	\cline{2-14}
&	Proposed (Rank 8)	&	251850	&	151013	&	267156	&	453856	&	199275	&	93701	&	165247	&	308654	&	303608	&	234658	&	409134	&	641950	\\	\cline{2-14}
&	Proposed (Rank 16)	&	309322	&	245463	&	439774	&	698158	&	269341	&	170374	&	302438	&	513257	&	385171	&	394167	&	678570	&	1001574	\\	\cline{2-14}
&	Proposed (Rank 28)	&	370101	&	337259	&	595617	&	933555	&	341876	&	266334	&	474772	&	752003	&	454738	&	520966	&	898310	&	1310081	\\	\cline{2-14}
&	Proposed (Rank 44)	&	434134	&	425824	&	737800	&	1151208	&	404438	&	384524	&	655959	&	1001388	&	507159	&	631820	&	1073918	&	1528251	\\	\cline{2-14}
&	Proposed (Rank 52)	&	459408	&	462759	&	794905	&	1220896	&	425346	&	432563	&	711515	&	1073548	&	524083	&	727147	&	1192872	&	1641494	\\	\cline{2-14}
&	Proposed (Rank 60)	&	481777	&	511679	&	855678	&	1289391	&	440675	&	457735	&	766217	&	1149042	&	535587	&	785724	&	1271113	&	1730093	\\	\cline{1-14}
\hline
\multirow{8}{*}{10} &	Dib E et al.	&	752723	&	1222920	&	1913852	&	4856266	&	777216	&	801876	&	1330787	&	3442540	&	782679	&	1345211	&	2155105	&	5205636	\\	\cline{2-14}
&	Proposed (Rank 4)	&	70222	&	49881	&	89829	&	166182	&	52813	&	29982	&	50755	&	103849	&	114218	&	79685	&	149995	&	246078	\\	\cline{2-14}
&	Proposed (Rank 8)	&	103595	&	83335	&	144000	&	247091	&	73658	&	44585	&	77578	&	148404	&	152447	&	128770	&	220654	&	346457	\\	\cline{2-14}
&	Proposed (Rank 16)	&	150921	&	135894	&	243209	&	385532	&	123883	&	87714	&	154861	&	265680	&	216590	&	228011	&	389982	&	561801	\\	\cline{2-14}
&	Proposed (Rank 28)	&	200848	&	192706	&	340303	&	523869	&	178811	&	140370	&	254368	&	402085	&	271876	&	309157	&	522290	&	743004	\\	\cline{2-14}
&	Proposed (Rank 44)	&	252927	&	249671	&	430367	&	652158	&	225984	&	223766	&	375786	&	553262	&	313974	&	379414	&	635701	&	880929	\\	\cline{2-14}
&	Proposed (Rank 52)	&	272641	&	273179	&	463991	&	690749	&	241537	&	245429	&	399996	&	582554	&	327708	&	442106	&	715530	&	955029	\\	\cline{2-14}
&	Proposed (Rank 60)	&	289867	&	299732	&	497046	&	726508	&	253575	&	272199	&	437409	&	627469	&	336198	&	477148	&	759510	&	1002492	\\	\cline{1-14}
\hline
\multirow{8}{*}{14}  &	Dib E et al.	&	480892	&	679589	&	1050351	&	2626996	&	521633	&	383952	&	638700	&	1648073	&	500828	&	742109	&	1187800	&	2801625	\\	\cline{2-14}
&	Proposed (Rank 4)	&	40679	&	29652	&	51278	&	98485	&	29001	&	15388	&	25291	&	56627	&	65006	&	43589	&	81743	&	136374	\\	\cline{2-14}
&	Proposed (Rank 8)	&	61694	&	44851	&	77249	&	140007	&	40184	&	23177	&	39845	&	80289	&	88848	&	66867	&	116614	&	189391	\\	\cline{2-14}
&	Proposed (Rank 16)	&	91688	&	75524	&	135330	&	219789	&	69269	&	46427	&	79815	&	142707	&	131125	&	127103	&	212521	&	310005	\\	\cline{2-14}
&	Proposed (Rank 28)	&	124841	&	108047	&	190712	&	296426	&	102889	&	75180	&	133133	&	214414	&	171095	&	175954	&	294598	&	419739	\\	\cline{2-14}
&	Proposed (Rank 44)	&	158055	&	140028	&	240595	&	365750	&	133740	&	116563	&	195948	&	288671	&	198596	&	215928	&	358409	&	496160	\\	\cline{2-14}
&	Proposed (Rank 52)	&	172129	&	153785	&	259051	&	385398	&	143679	&	127010	&	209194	&	303887	&	207304	&	252547	&	401949	&	532371	\\	\cline{2-14}
&	Proposed (Rank 60)	&	183237	&	167289	&	274776	&	401261	&	152195	&	136239	&	222340	&	319578	&	213372	&	275681	&	430690	&	564375	\\	\cline{1-14}
\hline
\multirow{8}{*}{20} &	Dib E et al.	&	213090	&	254203	&	381661	&	936738	&	237383	&	78987	&	150403	&	386715	&	220461	&	274582	&	434129	&	997828	\\	\cline{2-14}
&	Proposed (Rank 4)	&	20074	&	11978	&	20998	&	45282	&	13422	&	5936	&	9649	&	25904	&	30830	&	17895	&	33796	&	60029	\\	\cline{2-14}
&	Proposed (Rank 8)	&	29699	&	18356	&	30789	&	60782	&	18083	&	9076	&	14808	&	34361	&	41525	&	28044	&	48058	&	81843	\\	\cline{2-14}
&	Proposed (Rank 16)	&	43104	&	29966	&	52919	&	92344	&	29269	&	17566	&	29273	&	57232	&	60581	&	47269	&	78364	&	121263	\\	\cline{2-14}
&	Proposed (Rank 28)	&	58382	&	42027	&	73297	&	119860	&	42490	&	27054	&	49257	&	83411	&	79556	&	66517	&	110634	&	163393	\\	\cline{2-14}
&	Proposed (Rank 44)	&	73197	&	53595	&	90516	&	142227	&	56027	&	39443	&	64624	&	100884	&	93770	&	81035	&	133102	&	188734	\\	\cline{2-14}
&	Proposed (Rank 52)	&	78841	&	58178	&	96110	&	147314	&	60266	&	44362	&	71646	&	108650	&	98119	&	93685	&	147415	&	200453	\\	\cline{2-14}
&	Proposed (Rank 60)	&	83557	&	62421	&	100162	&	151331	&	63644	&	46378	&	73782	&	110035	&	100506	&	98744	&	154522	&	208763	\\	\cline{1-14}
\hline
\multirow{8}{*}{26} &	Dib E et al.	&	79458	&	76052	&	111845	&	267957	&	74881	&	8010	&	18414	&	52494	&	81501	&	76202	&	124228	&	276944	\\	\cline{2-14}
&	Proposed (Rank 4)	&	9863	&	5121	&	8570	&	23184	&	6554	&	2967	&	4983	&	13967	&	15128	&	7839	&	15117	&	28897	\\	\cline{2-14}
&	Proposed (Rank 8)	&	14088	&	8012	&	13369	&	33342	&	8505	&	4174	&	6644	&	17217	&	18704	&	11299	&	19552	&	35754	\\	\cline{2-14}
&	Proposed (Rank 16)	&	19741	&	11666	&	19789	&	38861	&	12568	&	7266	&	12305	&	24932	&	25352	&	15455	&	26860	&	44990	\\	\cline{2-14}
&	Proposed (Rank 28)	&	25182	&	15000	&	25496	&	46274	&	16190	&	9206	&	15165	&	29982	&	32582	&	21640	&	35738	&	57038	\\	\cline{2-14}
&	Proposed (Rank 44)	&	30210	&	17987	&	29681	&	50997	&	20472	&	10873	&	19081	&	34787	&	37997	&	30815	&	46664	&	68833	\\	\cline{2-14}
&	Proposed (Rank 52)	&	32119	&	19219	&	31116	&	52433	&	22260	&	11543	&	20113	&	36221	&	39422	&	27733	&	44203	&	65498	\\	\cline{2-14}
&	Proposed (Rank 60)	&	33288	&	19430	&	31422	&	52858	&	23179	&	11864	&	20961	&	37382	&	40662	&	28354	&	44886	&	66797	\\	\cline{1-14}
\hline
\multirow{8}{*}{38} &	Dib E et al.	&	13365	&	5614	&	8039	&	19199	&	9983	&	737	&	1149	&	2903	&	13141	&	3418	&	6028	&	13529	\\	\cline{2-14}
&	Proposed (Rank 4)	&	2303	&	1128	&	1700	&	5087	&	1466	&	787	&	1257	&	3905	&	2733	&	1701	&	2922	&	6111	\\	\cline{2-14}
&	Proposed (Rank 8)	&	3256	&	1399	&	2215	&	6353	&	1831	&	796	&	1249	&	3973	&	3418	&	1875	&	3098	&	6911	\\	\cline{2-14}
&	Proposed (Rank 16)	&	4087	&	1792	&	2941	&	6919	&	2306	&	919	&	1475	&	4642	&	3888	&	2140	&	3599	&	7383	\\	\cline{2-14}
&	Proposed (Rank 28)	&	5124	&	2025	&	3500	&	8088	&	2666	&	1017	&	1765	&	5011	&	4503	&	2291	&	4025	&	8008	\\	\cline{2-14}
&	Proposed (Rank 44)	&	5542	&	2349	&	3864	&	8427	&	2949	&	1157	&	1992	&	5248	&	5240	&	2339	&	3980	&	7964	\\	\cline{2-14}
&	Proposed (Rank 52)	&	5754	&	2458	&	3969	&	8508	&	3069	&	1177	&	2105	&	5587	&	5296	&	2432	&	4122	&	8436	\\	\cline{2-14}
&	Proposed (Rank 60)	&	5863	&	2536	&	4048	&	8389	&	3101	&	1149	&	1991	&	5345	&	5558	&	2435	&	4312	&	8577	\\	\cline{1-14}
\hline
\end{tabular} }
}
\end{table*}

\begin{table*}
\centering
\caption{\footnotesize Total number of bytes for each subset of Hierarchical-4 pattern using our proposed coding scheme and Dib et al. \cite{dib2019light}. }
\label{table_bytesH4}
\resizebox{6cm}{!}{
\makebox[\linewidth]{\begin{tabular}{|c|c|c|c|c|c|c|c|c|c|c|c|c|c|}
\hline
\multicolumn{2}{|c|}{} & \multicolumn{4}{|c|}{Bikes} & \multicolumn{4}{|c|}{Stone pillars outside}  & \multicolumn{4}{|c|}{Fountain \& Vincent 2} \\
\hline
QP & Scheme & Subset 1 & Subset 2 & Subset 3 & Subset 4 & Subset 1 & Subset 2 & Subset 3 & Subset 4 & Subset 1 & Subset 2 & Subset 3 & Subset 4\\
\hline
\multirow{8}{*}{2} &	Dib E et al.	&	1440724	&	1452789	&	5165455	&	14131746	&	1441216	&	1362428	&	4475726	&	12790584	&	1491971	&	1436217	&	4898061	&	13465870	\\	\cline{2-14}
&	Proposed (Rank 4)	&	369210	&	134008	&	323631	&	711432	&	319254	&	115673	&	265421	&	578809	&	427823	&	168575	&	540513	&	976300	\\	\cline{2-14}
&	Proposed (Rank 8)	&	449987	&	189225	&	491920	&	944449	&	365506	&	138944	&	354594	&	733985	&	497979	&	236676	&	758310	&	1293087	\\	\cline{2-14}
&	Proposed (Rank 16)	&	527356	&	284122	&	740612	&	1320068	&	461363	&	199211	&	550438	&	1042584	&	587856	&	308720	&	1036070	&	1653996	\\	\cline{2-14}
&	Proposed (Rank 28)	&	594826	&	330343	&	938636	&	1687787	&	542214	&	264184	&	790867	&	1374410	&	669761	&	373958	&	1307700	&	2043902	\\	\cline{2-14}
&	Proposed (Rank 44)	&	654269	&	396269	&	1161377	&	1988746	&	607986	&	334920	&	1018052	&	1718058	&	730933	&	513132	&	1660083	&	2511724	\\	\cline{2-14}
&	Proposed (Rank 52)	&	675815	&	414816	&	1235023	&	2107314	&	632929	&	353228	&	1118538	&	1877849	&	757197	&	551166	&	1824076	&	2674922	\\	\cline{2-14}
&	Proposed (Rank 60)	&	696291	&	488035	&	1460949	&	2399511	&	650310	&	381130	&	1185575	&	1980309	&	776640	&	561102	&	1783246	&	2694231	\\	\cline{1-14}
\hline
\multirow{8}{*}{6} &	Dib E et al.	&	1111819	&	1038264	&	3514681	&	9168841	&	1108750	&	960856	&	2990048	&	8187927	&	1165584	&	1034953	&	3260693	&	8419196	\\	\cline{2-14}
&	Proposed (Rank 4)	&	204317	&	62436	&	141660	&	311661	&	170346	&	51465	&	108216	&	241181	&	237931	&	96343	&	276107	&	473329	\\	\cline{2-14}
&	Proposed (Rank 8)	&	245169	&	101036	&	243053	&	467936	&	197249	&	71588	&	166252	&	344009	&	283215	&	137939	&	410716	&	663983	\\	\cline{2-14}
&	Proposed (Rank 16)	&	294934	&	153306	&	400324	&	701366	&	255205	&	110146	&	291160	&	544525	&	348866	&	195417	&	615257	&	934284	\\	\cline{2-14}
&	Proposed (Rank 28)	&	344636	&	195036	&	539529	&	919037	&	314895	&	162077	&	451320	&	755049	&	410858	&	245260	&	809494	&	1190873	\\	\cline{2-14}
&	Proposed (Rank 44)	&	395944	&	243143	&	691427	&	1120180	&	365865	&	208893	&	597212	&	959611	&	463453	&	324537	&	1030603	&	1465281	\\	\cline{2-14}
&	Proposed (Rank 52)	&	415335	&	260202	&	745264	&	1191375	&	388987	&	224727	&	666304	&	1052010	&	488418	&	354981	&	1146980	&	1579657	\\	\cline{2-14}
&	Proposed (Rank 60)	&	431383	&	300783	&	885964	&	1357140	&	399568	&	236506	&	701288	&	1100371	&	504481	&	383573	&	1159853	&	1615488	\\	\cline{1-14}
\hline
\multirow{8}{*}{10} &	Dib E et al.	&	775655	&	666938	&	2142004	&	5440908	&	775397	&	625993	&	1827285	&	4832009	&	827917	&	670702	&	1897787	&	4726968	\\	\cline{2-14}
&	Proposed (Rank 4)	&	62635	&	37584	&	75744	&	172831	&	50046	&	25982	&	50848	&	126123	&	100594	&	57683	&	146200	&	258515	\\	\cline{2-14}
&	Proposed (Rank 8)	&	95314	&	60742	&	135199	&	271180	&	69850	&	35405	&	79321	&	176440	&	135538	&	85910	&	228956	&	369946	\\	\cline{2-14}
&	Proposed (Rank 16)	&	137298	&	96352	&	230464	&	411368	&	113623	&	64943	&	154415	&	293640	&	185272	&	126344	&	364217	&	548547	\\	\cline{2-14}
&	Proposed (Rank 28)	&	179852	&	124355	&	318267	&	535924	&	160773	&	100419	&	257622	&	431745	&	235083	&	160710	&	494057	&	711798	\\	\cline{2-14}
&	Proposed (Rank 44)	&	222286	&	157319	&	416397	&	663668	&	199480	&	129589	&	348407	&	554301	&	277377	&	207598	&	626034	&	870952	\\	\cline{2-14}
&	Proposed (Rank 52)	&	237984	&	167861	&	448807	&	703844	&	213343	&	143137	&	394288	&	609943	&	293737	&	229906	&	711584	&	949928	\\	\cline{2-14}
&	Proposed (Rank 60)	&	252828	&	190614	&	532675	&	790999	&	224197	&	145857	&	405564	&	625081	&	305891	&	247286	&	714102	&	964923	\\	\cline{1-14}
\hline
\multirow{8}{*}{14}  &	Dib E et al.	&	500049	&	388083	&	1191431	&	2963845	&	513183	&	378425	&	1007183	&	2594115	&	541926	&	378184	&	988038	&	2383130	\\	\cline{2-14}
&	Proposed (Rank 4)	&	37151	&	21951	&	41475	&	104839	&	28674	&	16217	&	30015	&	74285	&	56057	&	35279	&	82651	&	149587	\\	\cline{2-14}
&	Proposed (Rank 8)	&	57510	&	37209	&	75972	&	161312	&	39181	&	21571	&	43882	&	103726	&	78222	&	52274	&	125153	&	213294	\\	\cline{2-14}
&	Proposed (Rank 16)	&	84762	&	59234	&	128341	&	243684	&	63968	&	37448	&	83762	&	167620	&	110915	&	78720	&	204995	&	319141	\\	\cline{2-14}
&	Proposed (Rank 28)	&	112659	&	77934	&	182347	&	317020	&	91763	&	55481	&	130123	&	240029	&	144823	&	100782	&	282460	&	414813	\\	\cline{2-14}
&	Proposed (Rank 44)	&	139924	&	97313	&	239016	&	389102	&	116872	&	76489	&	191692	&	313499	&	171103	&	121884	&	353503	&	497967	\\	\cline{2-14}
&	Proposed (Rank 52)	&	150018	&	104436	&	256212	&	409471	&	126814	&	82765	&	212476	&	335104	&	181934	&	144534	&	412929	&	555576	\\	\cline{2-14}
&	Proposed (Rank 60)	&	159098	&	115299	&	305078	&	457263	&	132712	&	88355	&	222964	&	347488	&	190145	&	145263	&	397671	&	546693	\\	\cline{1-14}
\hline
\multirow{8}{*}{20} &	Dib E et al.	&	220871	&	155877	&	442338	&	1069809	&	221668	&	144280	&	323891	&	808099	&	240428	&	132454	&	317605	&	742520	\\	\cline{2-14}
&	Proposed (Rank 4)	&	18693	&	9386	&	16817	&	49155	&	14681	&	6261	&	12388	&	38873	&	27684	&	17171	&	34590	&	69842	\\	\cline{2-14}
&	Proposed (Rank 8)	&	28822	&	17725	&	32732	&	79384	&	18981	&	10397	&	19570	&	51678	&	37767	&	23972	&	50490	&	96406	\\	\cline{2-14}
&	Proposed (Rank 16)	&	42123	&	27257	&	52383	&	112254	&	29375	&	17001	&	34307	&	76671	&	53174	&	35808	&	79916	&	136134	\\	\cline{2-14}
&	Proposed (Rank 28)	&	54836	&	35292	&	71902	&	136347	&	40267	&	21487	&	45165	&	93643	&	69187	&	45105	&	111945	&	176694	\\	\cline{2-14}
&	Proposed (Rank 44)	&	66687	&	42676	&	92133	&	161840	&	51066	&	29424	&	66927	&	119603	&	80368	&	54863	&	136769	&	200474	\\	\cline{2-14}
&	Proposed (Rank 52)	&	71351	&	45196	&	97830	&	166948	&	55199	&	30240	&	71069	&	121892	&	83943	&	59049	&	146132	&	205276	\\	\cline{2-14}
&	Proposed (Rank 60)	&	74991	&	47735	&	114919	&	183393	&	58046	&	33606	&	77786	&	128307	&	86632	&	61772	&	151950	&	212506	\\	\cline{1-14}
\hline
\multirow{8}{*}{26} &	Dib E et al.	&	80816	&	51560	&	127832	&	306862	&	70635	&	35903	&	66479	&	163981	&	88935	&	34759	&	71299	&	167645	\\	\cline{2-14}
&	Proposed (Rank 4)	&	9615	&	3901	&	7759	&	27704	&	7357	&	2136	&	4879	&	18330	&	13811	&	7502	&	13052	&	31445	\\	\cline{2-14}
&	Proposed (Rank 8)	&	14339	&	7033	&	13249	&	40246	&	9352	&	3875	&	7448	&	22526	&	17877	&	9379	&	19699	&	43266	\\	\cline{2-14}
&	Proposed (Rank 16)	&	19865	&	10767	&	19976	&	53272	&	13320	&	5723	&	10509	&	30017	&	23372	&	12724	&	25181	&	52259	\\	\cline{2-14}
&	Proposed (Rank 28)	&	25090	&	13451	&	25385	&	58593	&	16450	&	6534	&	14970	&	35654	&	29631	&	15747	&	34124	&	63501	\\	\cline{2-14}
&	Proposed (Rank 44)	&	29298	&	15253	&	30770	&	63818	&	19658	&	8361	&	16303	&	40251	&	33967	&	17384	&	37712	&	67354	\\	\cline{2-14}
&	Proposed (Rank 52)	&	30893	&	15025	&	29329	&	62347	&	20976	&	9012	&	19664	&	41450	&	35155	&	18862	&	42001	&	67561	\\	\cline{2-14}
&	Proposed (Rank 60)	&	31911	&	15795	&	32651	&	66267	&	22010	&	9312	&	20796	&	42488	&	35670	&	19892	&	43887	&	69423	\\	\cline{1-14}
\hline
\multirow{8}{*}{38} &	Dib E et al.	&	12808	&	5091	&	10029	&	22576	&	9452	&	2567	&	3519	&	6282	&	13878	&	1650	&	2798	&	5890	\\	\cline{2-14}
&	Proposed (Rank 4)	&	2070	&	697	&	1396	&	5099	&	1470	&	511	&	1120	&	4115	&	3161	&	1300	&	2589	&	7469	\\	\cline{2-14}
&	Proposed (Rank 8)	&	3312	&	1179	&	2248	&	7874	&	1817	&	604	&	1350	&	5095	&	3839	&	1278	&	2615	&	7893	\\	\cline{2-14}
&	Proposed (Rank 16)	&	4414	&	1452	&	2788	&	9105	&	2418	&	768	&	1562	&	5528	&	4534	&	1754	&	3159	&	8763	\\	\cline{2-14}
&	Proposed (Rank 28)	&	5149	&	1711	&	3140	&	9691	&	2826	&	777	&	1460	&	6012	&	5114	&	1616	&	3265	&	9168	\\	\cline{2-14}
&	Proposed (Rank 44)	&	5836	&	1842	&	3717	&	11701	&	3085	&	1173	&	2064	&	6325	&	5833	&	2084	&	4167	&	10721	\\	\cline{2-14}
&	Proposed (Rank 52)	&	6056	&	2071	&	3957	&	11951	&	3333	&	1056	&	1998	&	6683	&	5905	&	1891	&	4093	&	10013	\\	\cline{2-14}
&	Proposed (Rank 60)	&	6124	&	1844	&	3440	&	10214	&	3445	&	1113	&	2127	&	6741	&	5953	&	1987	&	3931	&	10061	\\	\cline{1-14}
\hline
\end{tabular} }
}
\end{table*}


\subsection{Experimental Settings and Implementation Details}\label{subsec41}

The proposed scheme is implemented on a single high-end HP OMEN X 15-DG0018TX system with 9th Gen i7-9750H, 16 GB RAM, RTX 2080 8 GB Graphics, and Windows 10 operating system. The multiplicative layers for each view subset of the four scanning orders are optimized using their corresponding convolutional neural network (CNN). The CNN contains twenty 2D convolutional layers stacked in a sequence. They have constant spatial size of tensors throughout and only the number of channels is varied. The first layer in each CNN corresponds to the tensor $\mathbf{L}$, which contains all the pixels of subset $L(u,v,s,t)$ being handled. $\mathbf{L}$ has number of channels equal to the number of viewpoints in respective subset and tensor $\mathbf{T}$ (that contained all the pixels of $T_{x}(u,v)$) has 3 channels corresponding to the 3 light field subset multiplicative layers. Intermediate convolutional layers comprise of 64 channels each. In Fig.~\ref{fig:workflow_encoding}, ‘ch’ refers to the channels in the convolutional layers. We train each CNN model for 25 epochs at a learning rate of 0.001 and a batch size of 15. The entire networks are implemented using the Python-based framework, Chainer (version 7.7.0).

The resultant output multiplicative layers for each subset of  $C_2$, $C_4$, $H_2$ and $H_4$ patterns are obtained from the trained CNN models. Fig.~\ref{fig:fountlayers} illustrates the three multiplicative layers produced for each view subset of Circular-2 and Circular-4 patterns in the \textit{Fountain \& Vincent 2} data. The three multiplicative layers of each subset of Hierarchical-2 and Hierarchical-4 are shown in Fig.~\ref{fig:bikeslayers} for the \textit{Bikes} data.  We rearranged the colour channels of these multiplicative layers as described in section~\ref{sec:bksvd} and then applied BK-SVD for ranks 4, 8, 16, 28, 44, 52 and 60. The approximated matrices are then arranged back into layers and compressed using HEVC (BLOCK II of COMPONENT I). We use quantization parameters 2, 6, 10, 14, 20, 26, and 38 to test both high and low bitrate cases of HEVC. The decoding and reconstruction of BK-SVD approximated subsets for all four patterns are then performed.

\begin{figure*}
\centering

\begin{subfigure}{0.2\textwidth}
  \includegraphics[width=\linewidth]{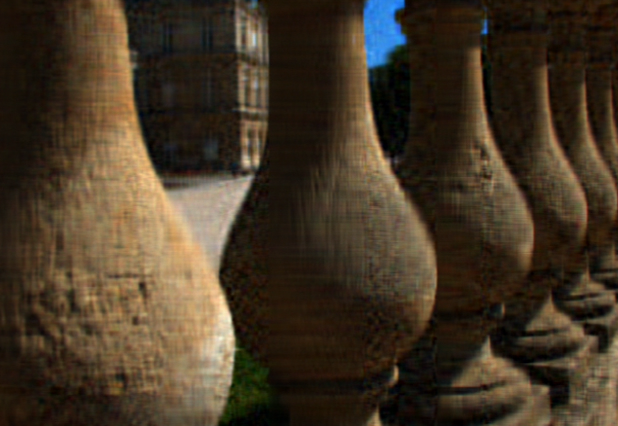}
\end{subfigure} 
\begin{subfigure}{0.2\textwidth}
  \includegraphics[width=\linewidth]{./Figures/c2_s_r28_q2.png}
\end{subfigure}
\begin{subfigure}{0.2\textwidth}
  \includegraphics[width=\linewidth]{./Figures/c2_s_r28_q2.png}
\end{subfigure} 

\begin{subfigure}{0.2\textwidth}
  \includegraphics[width=\linewidth]{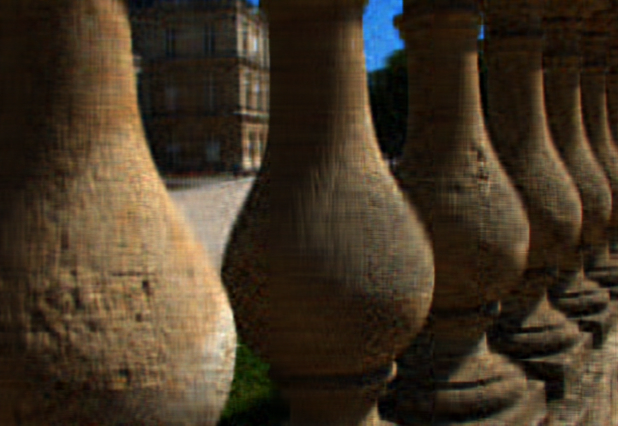}
\end{subfigure} 
\begin{subfigure}{0.2\textwidth}
  \includegraphics[width=\linewidth]{./Figures/c4_s_r28_q2.png}
\end{subfigure}
\begin{subfigure}{0.2\textwidth}
  \includegraphics[width=\linewidth]{./Figures/c4_s_r28_q2.png}
\end{subfigure} 

\begin{subfigure}{0.2\textwidth}
  \includegraphics[width=\linewidth]{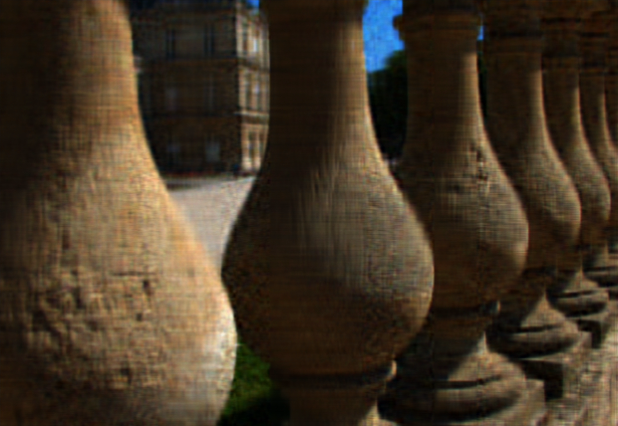}
\end{subfigure} 
\begin{subfigure}{0.2\textwidth}
  \includegraphics[width=\linewidth]{./Figures/h2_s_r28_q2.png}
\end{subfigure}
\begin{subfigure}{0.2\textwidth}
  \includegraphics[width=\linewidth]{./Figures/h2_s_r28_q2.png}
\end{subfigure}

\begin{subfigure}{0.2\textwidth}
  \includegraphics[width=\linewidth]{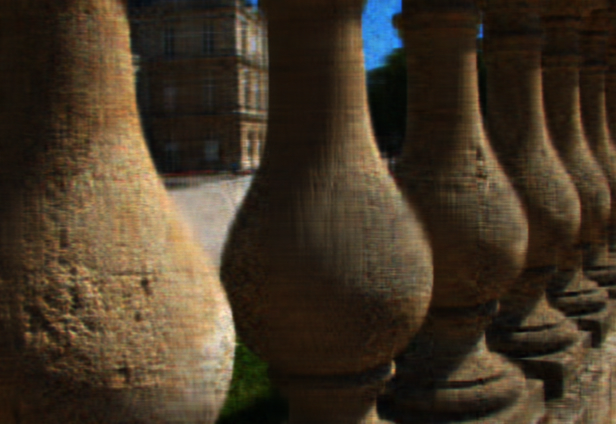}
    \caption{\footnotesize Rank 28}
\end{subfigure} 
\begin{subfigure}{0.2\textwidth}
  \includegraphics[width=\linewidth]{./Figures/h4_s_r28_q2.png}
  \caption{\footnotesize Rank 44}
\end{subfigure}
\begin{subfigure}{0.2\textwidth}
  \includegraphics[width=\linewidth]{./Figures/h4_s_r28_q2.png}
  \caption{\footnotesize Rank 60}
\end{subfigure} 
\caption{\footnotesize The central views of the reconstructed \textit{Stone pillars outside} data for ranks 28, 44 and  60, encoded with a quantization of 2. The The first, second, third and fourth rows show results of Circular-2, Circular-4, Hierarchical-2 and Hierarchical-4 respectively.}
\label{fig:stone_recon}
\end{figure*}

Low-rank approximated subsets are then utilized to form the FDL representation of light fields. The number of layers in the FDL method are fixed to $n = 30$. Views in approximated Subset 1 construct the initial FDL representation. The subsequent view subsets are predicted from this FDL representation and the residues iteratively refine the FDL representation. We used HEVC  to perform the encoding in COMPONENT II, choosing quantization parameters 2, 6, 10, 14, 20, 26 and 38. The final reconstructed light field central views at the end of COMPONENT III for the \textit{Stone pillars outside} data is illustrated in Fig.~\ref{fig:stone_recon} for the BK-SVD ranks 28, 44 and 60.


\begin{table*}
\centering
\caption{\footnotesize Bjontegaard percentage rate savings for the proposed compression scheme with respect to Dib et al. \cite{dib2019light}. Negative values represent gains.}
\label{table_bdpsnr}
\resizebox{6cm}{!}{
\makebox[\linewidth]{\begin{tabular}{|c|c|c|c|c|c|c|c|c|c|c|c|c|c|}
\hline
 &  & \multicolumn{2}{|c|}{Circular-2} & \multicolumn{2}{|c|}{Hierarchical-2} & \multicolumn{4}{|c|}{Circular-4} & \multicolumn{4}{|c|}{Hierarchical-4} \\
\hline
Scene & Rank & Subset 1 & Subset 2 & Subset 1 & Subset 2 & Subset 1 & Subset 2 & Subset 3 & Subset 4 & Subset 1 & Subset 2 & Subset 3 & Subset 4\\
\hline
\multirow{7}{*}{Bikes} &	4	&	-97.56541214	&	-99.29882593	&	-98.3907291	&	-99.54258115 &	-96.20857709	&	-99.34681784	&	-99.32842039	&	-99.34121406	&	-96.7351185	&	-99.00868427	&	-99.47500508	&	-99.34379191	\\	\cline{2-14}
&	8	&	-95.76892379	&	-99.03123267	&	-97.66246664	&	-99.29956835 &	-93.72289918	&	-98.65123258	&	-98.60472387	&	-98.82635815	&	-94.28513258	&	-97.74418081	&	-98.647811	&	-98.71977919	\\	\cline{2-14}
&	16	&	-93.17474979	&	-98.45339444	&	-96.63095962	&	-98.95497164 &	-89.76525824	&	-97.16858635	&	-96.92654865	&	-98.03975782	&	-90.71898512	&	-95.38280704	&	-97.04688734	&	-97.88951207	\\	\cline{2-14}
&	28	&	-90.3477664	&	-98.11361299	&	-95.58497187	&	-98.67266268 &	-84.94581006	&	-95.34857594	&	-94.83394299	&	-97.03298825	&	-86.86256935	&	-92.89572507	&	-95.15379369	&	-97.0052932	\\	\cline{2-14}
&	44	&	-87.34955344	&	-97.75859383	&	-94.60404741	&	-98.07437089 &	-79.85797265	&	-92.77852898	&	-92.20674636	&	-95.85782231	&	-82.89202874	&	-90.05487069	&	-92.67313576	&	-96.06675045	\\	\cline{2-14}
&	52	&	-86.18089698	&	-97.59830221	&	-94.16664723	&	-97.93405439 &	-77.60680642	&	-91.69706112	&	-91.16454361	&	-95.4561184	&	-81.31504874	&	-89.00949178	&	-91.89758467	&	-95.79899808	\\	\cline{2-14}
&	60	&	-85.37903466	&	-97.4382535	&	-93.89176465	&	-97.88383496 &	-75.79473857	&	-90.6698963	&	-90.38054322	&	-95.1466604	&	-80.07701534	&	-87.21412008	&	-90.0680445	&	-95.07426992	\\	\cline{1-14}
\hline

\multirow{7}{*}{Stone pillars outside } &	4	&	-98.28738998	&	-99.5671618	&	-98.54643236	&	-99.69494577 &	-96.70390985	&	-99.63514874	&	-99.63502461	&	-99.59656973	&	-96.57281486	&	-99.34377591	&	-99.50760509	&	-99.37042389	\\	\cline{2-14}
&	8	&	-97.48146218	&	-99.45696678	&	-97.88804729	&	-99.65675017 &	-96.83115691	&	-99.35667758	&	-99.32191491	&	-99.35202452	&	-96.76795633	&	-99.10596584	&	-98.99132784	&	-99.12125107	\\	\cline{2-14}
&	16	&	-95.3611157	&	-98.79333481	&	-96.93436888	&	-99.10493755 &	-93.99292616	&	-98.16597465	&	-98.2962672	&	-98.53366947	&	-94.36508582	&	-98.03801948	&	-98.47550354	&	-98.38338342	\\	\cline{2-14}
&	28	&	-93.25557142	&	-98.75966092	&	-95.85758052	&	-98.75257079 &	-90.4631812	&	-96.53677355	&	-96.49717131	&	-97.50938121	&	-91.45018488	&	-97.04731002	&	-97.62269234	&	-98.45732096	\\	\cline{2-14}
&	44	&	-91.33970709	&	-98.53538189	&	-94.86352824	&	-98.5749095	&	-86.68637432	&	-93.88697432	&	-94.29482711	&	-97.03694131	&	-88.48088951	&	-95.14677527	&	-95.98370386	&	-97.95248612 \\	\cline{2-14}
&	52	&	-91.07697181	&	-98.45815322	&	-94.57758068	&	-98.46705075 &	-85.22044971	&	-93.13489515	&	-93.67714426	&	-96.75798999	&	-87.20785431	&	-94.70593446	&	-95.65455146	&	-97.72227925	\\	\cline{2-14}
&	60	&	-90.76161372	&	-98.35636369	&	-94.30087553	&	-98.40819422 &	-84.15256024	&	-93.01769387	&	-93.33095286	&	-96.66548892	&	-86.25854702	&	-94.27613508	&	-95.06018928	&	-97.5857605	\\	\cline{1-14}
\hline

\multirow{7}{*}{Fountain \& Vincent 2} &	4	&	-96.27656828	&	-99.26603902	&	-96.9530291	&	-99.57099646 &	-93.24324684	&	-98.55291445	&	-98.36932301	&	-98.98906935	&	-94.92713556	&	-97.50272257	&	-98.13359141	&	-98.93827929	\\	\cline{2-14}
&	8	&	-94.48002919	&	-98.96179277	&	-96.34841383	&	-99.08720888 &	-90.2813932	&	-97.37199047	&	-97.20967603	&	-98.62957034	&	-92.52170343	&	-95.77161689	&	-96.4265349	&	-98.44379923	\\	\cline{2-14}
&	16	&	-91.14745649	&	-99.02010069	&	-95.17447177	&	-99.23270338 &	-84.41060243	&	-94.13934209	&	-93.86215798	&	-97.62140921	&	-88.53676951	&	-92.8003062	&	-94.58597524	&	-97.81833936	 \\	\cline{2-14}
&	28	&	-87.71507969	&	-98.92810522	&	-93.90335982	&	-98.97503776 &	-78.21793701	&	-92.83126167	&	-91.91796171	&	-96.58176205	&	-83.95868385	&	-91.25156494	&	-92.28424221	&	-96.94553427	\\	\cline{2-14}
&	44	&	-85.20018338	&	-98.74257926	&	-92.71888896	&	-98.94441882 &	-73.07216196	&	-91.43574166	&	-90.58850443	&	-95.97145607	&	-80.11693215	&	-86.79417239	&	-88.96051247	&	-95.75589133	\\	\cline{2-14}
&	52	&	-84.44629524	&	-98.77812886	&	-92.50450918	&	-98.83289017 &	-71.5870472	&	-89.37972401	&	-88.61773338	&	-95.48946614	&	-78.80228046	&	-85.31225663	&	-87.39959326	&	-95.3181428\\	\cline{2-14}
&	60	&	-83.73495541	&	-98.74446465	&	-92.34241435	&	-98.90331735 &	-70.27253249	&	-88.33184967	&	-87.93389356	&	-95.17732486	&	-77.72435271	&	-85.15958083	&	-87.64733413	&	-95.43459988	\\	\cline{1-14}

\hline

\end{tabular} }
}
\end{table*}

\subsection{Results and Comparative Analysis}

We compare the proposed scheme with Dib et al.~\cite{dib2019light} light field coding algorithm. The proposed coding scheme outperforms by large margins. The total number of bytes taken by our approach is comparatively far lesser than the work by Dib et al.~\cite{dib2019light} for all ranks and QPs. The corresponding results are depicted in Table~\ref{table_bytesC2} for $C_2$, Table~\ref{table_bytesH2} for $H_2$, Table~\ref{table_bytesC4} for $C_4$ and Table~\ref{table_bytesH4} for $H_4$ patterns. The bitrate vs YUV-PSNR graphs of three datasets in $C_2$, $C_4$, $H_2$ and $H_4$ configurations are illustrated in Fig.~\ref{fig:graphs}. For all four scanning patterns, the proposed scheme has significantly better rate-distortion results considering both subset-wise light field and entire light field. 

Further, we analyze  bitrate reduction (BD-rate) of the proposed scheme with respect to Dib et al.~\cite{dib2019light} using Bjontgaard metric~\cite{bjontegaard2001calculation}. The average percent difference in rate change is estimated over a range of QPs for seven chosen ranks. A comparison of the percentage of bitrate savings of our proposed coding scheme with respect to the anchor method for three chosen light field datasets is shown in Table~\ref{table_bdpsnr}. For $C_2$ pattern, the proposed scheme achieves $94.53\%$, $96.39\%$, and $93.96\%$ bitrate reduction compared to Dib et al.~\cite{dib2019light} for light fields \textit{Bikes}, \textit{Stone pillars outside}, and \textit{Fountain \& Vincent 2} respectively. For pattern $H_2$, the proposed scheme achieves $97.23\%$, $97.54\%$, and $96.67\%$ bitrate reduction compared to Dib et al.~\cite{dib2019light} for the light fields \textit{Bikes}, \textit{Stone pillars outside}, and \textit{Fountain \& Vincent 2} respectively. Lastly, the $C_4$ and $H_4$ patterns have $93.09\%$, $95.29\%$, $90.71\%$, and $93.18\%$, $96.02\%$, $91.25\%$ bitrate reduction respectively in the proposed scheme over Dib et al.~\cite{dib2019light} for \textit{Bikes}, \textit{Stone pillars outside}, and \textit{Fountain \& Vincent 2}.


\section{Conclusion}\label{con}
We have proposed a novel hierarchical hybrid coding scheme for light fields based on transmittance patterns of low-rank multiplicative layers and Fourier Disparity Layers. Typical pseudo sequence based light field compression schemes \cite{RwPsvRef1_liu2016pseudo,RwPsvRef3_ahmad2017interpreting,RwPsvRef4_ahmad2019computationally,RwPsvRef5_gu2019high} do not efficiently consider the similarities between horizontal and vertical views of a light field. Our proposed scheme not only exploits the spatial correlation in multiplicative layers of the subset views for varying low ranks, but also removes the temporal intra and inter-layer redundancies in the low-rank approximated representation of view subsets. The approximated light field is further compressed by eliminating intrinsic similarities among neighboring views of Circular-2, Circular-4, Hierarchical-2 and Hierarchical-4 patterns using Fourier Disparity Layers representation. This integrated compression achieves excellent bitrate reductions without compromising the quality of the reconstructed light field.

Our scheme offers flexibility to cover a range of multiple bitrates using just a few trained CNNs to obtain a layered representation of the light field subsets. This critical feature sets our proposed scheme apart from other existing light field coding methods, which usually train a system (or network) to support only specific bitrates during the compression. Besides, existing coding approaches are not explicitly designed to target layered displays and are usually only classified to work for lenslet-based formats or sub-aperture images based pseudo-sequence light field representation. The proposed flexible coding scheme can support not just multi-layered light field displays, but is also adaptable to table-top~\cite{maruyama2018implementation} or other variety of autostereoscopic displays.

In our future work, we plan to extend the proposed idea to light field displays with more than three light attenuating layers. Proof-of-concept experiments with our scheme also pave the way to form a more profound understanding in the rank-analysis of a light-field using other mathematically valid tensor-based models~\cite{wetzstein2012tensor,kobayashi2017focal} and coded mask cameras~\cite{maruyama2019coded}. We also aim to verify our scheme with physical light field display hardware. Further, we wish to adapt the proposed scheme with display device availability and optimize the bandwidth for a target bitrate. This would enable deploying the concepts of layered displays on different display platforms that deliver 3D contents with limited hardware resources and thus best meet the viewers’ preferences for depth impression and visual comfort.

\section*{Author contributions}
\textbf{Joshitha Ravishankar:} Methodology, Software, Validation, Formal analysis, Investigation, Data Curation, Writing- Original Draft, Visualization. \textbf{Mansi Sharma:} Conceptualization, Methodology, Software, Validation, Formal analysis, Investigation, Resources, Writing- Original Draft, Supervision, Project administration.


\section*{Acknowledgement}
The scientific efforts leading to the results reported in this paper have been carried out under the supervision of Dr. Mansi Sharma, INSPIRE Hosted Faculty, IIT Madras. This work has been supported, in part, by the Department of Science and Technology, Government of India project \textit{``Tools and Processes for Multi-view 3D Display Technologies''}, DST/INSPIRE/04/2017/001853.

We would like to thank Nikitha Varma Sunchu and P Sai Shankar Pavan Srinivas for their help in running the codes for experiments. 


\section*{Declaration of interests}
The authors declare that they have no known competing financial interests or personal relationships that could have appeared to influence the work reported in this paper.


\bibliography{root}

\begin{thebibliography}{10}
\expandafter\ifx\csname url\endcsname\relax
  \def\url#1{\texttt{#1}}\fi
\expandafter\ifx\csname urlprefix\endcsname\relax\def\urlprefix{URL }\fi
\expandafter\ifx\csname href\endcsname\relax
  \def\href#1#2{#2} \def\path#1{#1}\fi

\bibitem{surman2014towards}
P.~Surman, X.~W. Sun, Towards the reality of 3d imaging and display, in: 2014
  3DTV-Conference: The True Vision-Capture, Transmission and Display of 3D
  Video (3DTV-CON), IEEE, 2014, pp. 1--4.

\bibitem{balogh2007holovizio}
T.~Balogh, P.~T. Kov{\'a}cs, A.~Barsi, Holovizio 3d display system, in: 2007
  3DTV Conference, IEEE, 2007, pp. 1--4.

\bibitem{pseudo3d}
Y.~Kubota, N.~Umezu, Pseudo-3d display based on head tracking for viewpoint
  image generation, in: 2019 IEEE 8th Global Conference on Consumer Electronics
  (GCCE), 2019, pp. 158--161.
\newblock \href {http://dx.doi.org/10.1109/GCCE46687.2019.9015351}
  {\path{doi:10.1109/GCCE46687.2019.9015351}}.

\bibitem{glassesfree3d}
P.~Surman, X.~Zhang, W.~Song, X.~Xia, S.~Wang, Y.~Zheng, Glasses-free 3-d and
  augmented reality display advances: From theory to implementation, IEEE
  MultiMedia 27~(1) (2020) 17--26.
\newblock \href {http://dx.doi.org/10.1109/MMUL.2019.2948334}
  {\path{doi:10.1109/MMUL.2019.2948334}}.

\bibitem{li2020light}
T.~Li, Q.~Huang, S.~Alfaro, A.~Supikov, J.~Ratcliff, G.~Grover, R.~Azuma,
  Light-field displays: a view-dependent approach, in: ACM SIGGRAPH 2020 ET,
  2020, pp. 1--2.

\bibitem{wetzstein2012tensor}
G.~Wetzstein, D.~R. Lanman, M.~W. Hirsch, R.~Raskar, Tensor displays:
  compressive light field synthesis using multilayer displays with directional
  backlighting.

\bibitem{sharma2016novel}
M.~Sharma, S.~Chaudhury, B.~Lall, A novel hybrid kinect-variety-based
  high-quality multiview rendering scheme for glass-free 3d displays, IEEE
  TCSVT 27~(10) (2016) 2098--2117.

\bibitem{RwlensletRef1_li2014efficient}
Y.~Li, M.~Sj{\"o}str{\"o}m, R.~Olsson, U.~Jennehag, Efficient intra prediction
  scheme for light field image compression, in: 2014 IEEE International
  conference on acoustics, speech and signal processing (ICASSP), IEEE, 2014,
  pp. 539--543.

\bibitem{RwlensletRef2_perra2016high}
C.~Perra, P.~Assuncao, High efficiency coding of light field images based on
  tiling and pseudo-temporal data arrangement, in: 2016 IEEE International
  Conference on Multimedia \& Expo Workshops (ICMEW), IEEE, 2016, pp. 1--4.

\bibitem{RwlensletRef3_li2016compression}
Y.~Li, R.~Olsson, M.~Sj{\"o}str{\"o}m, Compression of unfocused plenoptic
  images using a displacement intra prediction, in: 2016 IEEE ICMEW, IEEE,
  2016, pp. 1--4.

\bibitem{RwlensletRef4_monteiro2017light}
R.~J. Monteiro, P.~J. Nunes, N.~M. Rodrigues, S.~M. Faria, Light field image
  coding using high-order intrablock prediction, IEEE JSTSP 11~(7) (2017)
  1120--1131.

\bibitem{RwlensletRef5_liu2019content}
D.~Liu, P.~An, R.~Ma, W.~Zhan, X.~Huang, A.~A. Yahya, Content-based light field
  image compression method with gaussian process regression, IEEE TM 22~(4)
  (2019) 846--859.

\bibitem{RwDispRef2_jiang2017light}
X.~Jiang, M.~Le~Pendu, R.~A. Farrugia, C.~Guillemot, Light field compression
  with homography-based low-rank approximation, IEEE JSTSP 11~(7) (2017)
  1132--1145.

\bibitem{RwDispRef1_zhao2017light}
S.~Zhao, Z.~Chen, Light field image coding via linear approximation prior, in:
  IEEE ICIP, IEEE, 2017, pp. 4562--4566.

\bibitem{RwEpiRef2_ahmad2020shearlet}
W.~Ahmad, S.~Vagharshakyan, M.~Sj{\"o}str{\"o}m, A.~Gotchev, R.~Bregovic,
  R.~Olsson, Shearlet transform-based light field compression under low
  bitrates, IEEE TIP 29 (2020) 4269--4280.

\bibitem{RwEpiRef3_chen2020light}
Y.~Chen, P.~An, X.~Huang, C.~Yang, D.~Liu, Q.~Wu, Light field compression using
  global multiplane representation and two-step prediction, IEEE SPL 27 (2020)
  1135--1139.

\bibitem{RwCbRef1_hu2020adaptive}
X.~Hu, J.~Shan, Y.~Liu, L.~Zhang, S.~Shirmohammadi, An adaptive two-layer light
  field compression scheme using gnn-based reconstruction, ACM TOMM 16~(2s)
  (2020) 1--23.

\bibitem{RwDispRef3_dib2020local}
E.~Dib, M.~Le~Pendu, X.~Jiang, C.~Guillemot, Local low rank approximation with
  a parametric disparity model for light field compression, IEEE TIP 29 (2020)
  9641--9653.

\bibitem{RwEpiRef1_vagharshakyan2017light}
S.~Vagharshakyan, R.~Bregovic, A.~Gotchev, Light field reconstruction using
  shearlet transform, IEEE transactions on pattern analysis and machine
  intelligence 40~(1) (2017) 133--147.

\bibitem{RwVsRef3_huang2019light}
X.~Huang, P.~An, F.~Cao, D.~Liu, Q.~Wu, Light-field compression using a pair of
  steps and depth estimation, Optics express 27~(3) (2019) 3557--3573.

\bibitem{RwVsRef4_heriard2019light}
B.~H{\'e}riard-Dubreuil, I.~Viola, T.~Ebrahimi, Light field compression using
  translation-assisted view estimation, in: 2019 PCS, IEEE, 2019, pp. 1--5.

\bibitem{RwDeepRef6_schiopu2019deep}
I.~Schiopu, A.~Munteanu, Deep-learning-based macro-pixel synthesis and lossless
  coding of light field images, APSIPA TSIP 8.

\bibitem{RwDeepRef8_liu2021view}
D.~Liu, X.~Huang, W.~Zhan, L.~Ai, X.~Zheng, S.~Cheng, View synthesis-based
  light field image compression using a generative adversarial network,
  Information Sciences 545 (2021) 118--131.

\bibitem{RwPsvRef1_liu2016pseudo}
D.~Liu, L.~Wang, L.~Li, Z.~Xiong, F.~Wu, W.~Zeng, Pseudo-sequence-based light
  field image compression, in: IEEE ICMEW, IEEE, 2016, pp. 1--4.

\bibitem{RwPsvRef3_ahmad2017interpreting}
W.~Ahmad, R.~Olsson, M.~Sj{\"o}str{\"o}m, Interpreting plenoptic images as
  multi-view sequences for improved compression, in: IEEE ICIP, IEEE, 2017, pp.
  4557--4561.

\bibitem{RwPsvRef4_ahmad2019computationally}
W.~Ahmad, M.~Ghafoor, S.~A. Tariq, A.~Hassan, M.~Sj{\"o}str{\"o}m, R.~Olsson,
  Computationally efficient light field image compression using a multiview
  hevc framework, IEEE access 7 (2019) 143002--143014.

\bibitem{RwPsvRef5_gu2019high}
J.~Gu, B.~Guo, J.~Wen, High efficiency light field compression via virtual
  reference and hierarchical mv-hevc, in: IEEE (ICME), IEEE, 2019, pp.
  344--349.

\bibitem{ravishankar2021flexible}
J.~Ravishankar, M.~Sharma, P.~Gopalakrishnan, A flexible coding scheme based on
  block krylov subspace approximation for light field displays with stacked
  multiplicative layers, Sensors 21~(13) (2021) 4574.

\bibitem{maruyama2020comparison}
K.~Maruyama, K.~Takahashi, T.~Fujii, Comparison of layer operations and
  optimization methods for light field display, IEEE Access 8 (2020)
  38767--38775.

\bibitem{le2019fourier}
M.~Le~Pendu, C.~Guillemot, A.~Smolic, A fourier disparity layer representation
  for light fields, IEEE TIP 28~(11) (2019) 5740--5753.

\bibitem{RwDeepRef3_bakir2018light}
N.~Bakir, W.~Hamidouche, O.~D{\'e}forges, K.~Samrouth, M.~Khalil, Light field
  image compression based on convolutional neural networks and linear
  approximation, in: 2018 25th IEEE International Conference on Image
  Processing (ICIP), IEEE, 2018, pp. 1128--1132.

\bibitem{RwDeepRef4_zhao2018light}
Z.~Zhao, S.~Wang, C.~Jia, X.~Zhang, S.~Ma, J.~Yang, Light field image
  compression based on deep learning, in: 2018 IEEE International Conference on
  Multimedia and Expo (ICME), IEEE, 2018, pp. 1--6.

\bibitem{RwDeepRef5_wang2019region}
B.~Wang, Q.~Peng, E.~Wang, K.~Han, W.~Xiang, Region-of-interest compression and
  view synthesis for light field video streaming, IEEE Access 7 (2019)
  41183--41192.

\bibitem{RwDeepRef7_jia2018light}
C.~Jia, X.~Zhang, S.~Wang, S.~Wang, S.~Ma, Light field image compression using
  generative adversarial network-based view synthesis, IEEE Journal on Emerging
  and Selected Topics in Circuits and Systems 9~(1) (2018) 177--189.

\bibitem{musco2015randomized}
C.~Musco, C.~Musco, Randomized block krylov methods for stronger and faster
  approximate singular value decomposition, arXiv preprint arXiv:1504.05477.

\bibitem{sullivan2012overview}
G.~J. Sullivan, J.-R. Ohm, W.-J. Han, T.~Wiegand, Overview of the high
  efficiency video coding (hevc) standard, IEEE TCSVT 22~(12) (2012)
  1649--1668.

\bibitem{dib2019light}
E.~Dib, M.~Le~Pendu, C.~Guillemot, Light field compression using fourier
  disparity layers, in: IEEE ICIP, IEEE, 2019, pp. 3751--3755.

\bibitem{DBLP:journals/corr/abs-2104-09378}
J.~Ravishankar, M.~Sharma, \href{https://arxiv.org/abs/2104.09378}{A
  hierarchical coding scheme for glasses-free 3d displays based on scalable
  hybrid layered representation of real-world light fields}, CoRR
  abs/2104.09378.
\newblock \href {http://arxiv.org/abs/2104.09378} {\path{arXiv:2104.09378}}.
\newline\urlprefix\url{https://arxiv.org/abs/2104.09378}

\bibitem{levoy1996light}
M.~Levoy, P.~Hanrahan, Light field rendering, in: Proceedings of the 23rd
  annual conference on Computer graphics and interactive techniques, 1996, pp.
  31--42.

\bibitem{gortler1996lumigraph}
S.~J. Gortler, R.~Grzeszczuk, R.~Szeliski, M.~F. Cohen, The lumigraph, in:
  Proceedings of the 23rd annual conference on Computer graphics and
  interactive techniques, 1996, pp. 43--54.

\bibitem{jpeg_pennebaker1992jpeg}
W.~B. Pennebaker, J.~L. Mitchell, JPEG: Still image data compression standard,
  Springer Science \& Business Media, 1992.

\bibitem{hevc_sullivan2012overview}
G.~J. Sullivan, J.-R. Ohm, W.-J. Han, T.~Wiegand, Overview of the high
  efficiency video coding (hevc) standard, IEEE Transactions on circuits and
  systems for video technology 22~(12) (2012) 1649--1668.

\bibitem{RwPsvRef2_li2017pseudo}
L.~Li, Z.~Li, B.~Li, D.~Liu, H.~Li, Pseudo-sequence-based 2-d hierarchical
  coding structure for light-field image compression, IEEE Journal of Selected
  Topics in Signal Processing 11~(7) (2017) 1107--1119.

\bibitem{RwVsRef1_senoh2018efficient}
T.~Senoh, K.~Yamamoto, N.~Tetsutani, H.~Yasuda, Efficient light field image
  coding with depth estimation and view synthesis, in: 2018 26th European
  Signal Processing Conference (EUSIPCO), IEEE, 2018, pp. 1840--1844.

\bibitem{RwVsRef2_huang2018view}
X.~Huang, P.~An, L.~Shan, R.~Ma, L.~Shen, View synthesis for light field coding
  using depth estimation, in: 2018 IEEE International Conference on Multimedia
  and Expo (ICME), IEEE, 2018, pp. 1--6.

\bibitem{mitdragon}
G.~Wetzstein, Synthetic light field archive - mit media lab,
  \url{https://web.media.mit.edu/~gordonw/SyntheticLightFields/}, accessed:
  01-04-2021.

\bibitem{gu1996efficient}
M.~Gu, S.~C. Eisenstat, Efficient algorithms for computing a strong
  rank-revealing qr factorization, SIAM Journal on Scientific Computing 17~(4)
  (1996) 848--869.

\bibitem{le2020hierarchical}
M.~Le~Pendu, C.~Ozcinar, A.~Smolic, Hierarchical fourier disparity layer
  transmission for light field streaming, in: 2020 IEEE International
  Conference on Image Processing (ICIP), IEEE, 2020, pp. 2606--2610.

\bibitem{le2020high}
M.~Le~Pendu, A.~Smolic, High resolution light field recovery with fourier
  disparity layer completion, demosaicing, and super-resolution, in: 2020 IEEE
  International Conference on Computational Photography (ICCP), IEEE, 2020, pp.
  1--12.

\bibitem{rerabek2016new}
M.~Rerabek, T.~Ebrahimi, New light field image dataset, in: 8th International
  Conference on QoMEX, 2016.

\bibitem{dansereau2013decoding}
D.~G. Dansereau, O.~Pizarro, S.~B. Williams, Decoding, calibration and
  rectification for lenselet-based plenoptic cameras, in: Proceedings of the
  IEEE conference on computer vision and pattern recognition, 2013, pp.
  1027--1034.

\bibitem{bjontegaard2001calculation}
G.~Bjontegaard, Calculation of average psnr differences between rd-curves,
  VCEG-M33.

\bibitem{maruyama2018implementation}
K.~Maruyama, H.~Kojima, K.~Takahashi, T.~Fujii, Implementation of table-top
  light-field display, in: IDW, 2018.

\bibitem{kobayashi2017focal}
Y.~Kobayashi, K.~Takahashi, T.~Fujii, From focal stacks to tensor display: A
  method for light field visualization without multi-view images, in: IEEE
  ICASSP, IEEE, 2017, pp. 2007--2011.

\bibitem{maruyama2019coded}
K.~Maruyama, Y.~Inagaki, K.~Takahashi, T.~Fujii, H.~Nagahara, A 3-d display
  pipeline from coded-aperture camera to tensor light-field display through
  cnn, in: IEEE ICIP, IEEE, 2019, pp. 1064--1068.

\end{thebibliography}

\end{document}